\documentclass[12pt,preprint]{aastex}

\shorttitle{Two Impostors }
\shortauthors{Humphreys et al. }

\begin{document}

\title{A Tale of Two Impostors: SN2002kg and SN1954J in NGC 2403\altaffilmark{1} }

\author{
Roberta M.. Humphreys\altaffilmark{2}, 
Kris Davidson\altaffilmark{2},
Schuyler D. Van Dyk\altaffilmark{3}
and
Michael S. Gordon\altaffilmark{2},  
}

\altaffiltext{1}  
{Based  on observations  with the Multiple Mirror Telescope, a joint facility of
the Smithsonian Institution and the University of Arizona, and on observations obtained 
with the Large Binocular Telescope (LBT), an international collaboration among 
institutions in the United States, Italy and Germany. LBT Corporation partners are: 
The University of
 Arizona on behalf of the Arizona university system; Istituto Nazionale di
 Astrofisica, Italy; LBT Beteiligungsgesellschaft, Germany, representing the
 Max-Planck Society, the Astrophysical Institute Potsdam, and Heidelberg
 University; The Ohio State University, and The Research Corporation, on
 behalf of The University of Notre Dame, University of Minnesota and
 University of Virginia. 
} 

\altaffiltext{2}
{Minnesota Institute for Astrophysics, 116 Church St SE, University of Minnesota
, Minneapolis, MN 55455; roberta@umn.edu} 

\altaffiltext{3}
{Caltech/IPAC, Mailcode 100-22, Pasadena, CA 91125}

\begin{abstract}
We describe new results on two supernova impostors in NGC 2403,  
SN 1954J(V12) and SN 2002kg(V37). For the famous object 
SN 1954J we combine four critical observations:  
its current SED, its H$\alpha$ emission line profile, 
 the \ion{Ca}{2} triplet in absorption in its red spectrum, 
and the brightness compared 
to its pre-event state.  Together these strongly suggest that 
the survivor is now a hot supergiant with $T \sim 20000$ K, a 
dense wind, substantial circumstellar extinction, and a  
G-type supergiant companion. The hot star progenitor of V12's giant eruption 
was likely  in the post-red supergiant stage and had already 
shed a lot of mass.  
V37 is a classical LBV/S Dor variable.  Our photometry and spectra observed  
during and after its eruption show that its outburst was an apparent 
transit on the HR Diagram due to enhanced mass loss and the formation of 
a cooler, dense wind. V37  is an evolved hot supergiant at 
$\approx$ 10$^{6}$ L$_{\odot}$ with a probable initial mass of 
60 -80 M$_{\odot}$.   
\end{abstract} 


%%%%%%

\section{Introduction}  %%% ===-=== %%%  
\label{sec:intro}  

Massive star ``eruptions'' basically fall into two categories:  normal 
Luminous Blue Variable(LBV)/S Doradus variability and giant eruptions with super-Eddington
luminosities and massive radiation-driven outflows \citep{HD94,Owocki,kd2016}. 
In a giant eruption the total luminosity substantially increases.
The  energetics of the giant eruptions, the duration, and what we observe are
quite different from  LBV/S Dor variability.  The so-called LBV eruption 
is due to an increased mass loss rate which causes the
wind to become dense and opaque with a large, cooler pseudo-photosphere
and no substantial increase in luminosity in most cases. 
This enhanced mass loss state can last for several years.
The origin  of the instability,  their progenitors, and  
evolutionary state of the giant eruptins and the classical LBVs 
may also be different.

The numerous transient surveys  are finding an increasing number of 
what appear to be non-terminal giant eruptions, apparently from evolved 
very massive stars. Many of these giant eruptions are spectroscopically similar to the Type IIn
supernovae and thus receive a SN designation, but are later recognized as
sub-luminous or their spectra and light curves do not develop like true
supernovae. Consequently, they are often referred to as ``supernova impostors''
( see \citet{VanDyk12b} for a review). This is a  diverse group of objects with a range of 
luminosities and possible progenitors. A few impostors appear to be normal LBV/S Dor
variables in the eruptive or maximum light state, while most are giant eruptions possibly similar to $\eta$ Car \citep{HD94}. Of course the true test of an impostor is that it survives the giant eruption. Three of the four historical
examples, $\eta$ Car, P Cyg, and SN1954J \citep{RMH99,VanDyk05b}, survived               their giant eruptions while the fourth, SN1961V, is still controversial \citep{CSK11,VanDyk12a}.

Here we present recent spectra and photometry for two supernova impostors in the nearby 
spiral NGC2403: SN1954J(V12) and SN2002kg(V37). Both received supernova
designations, but instead are examples of non-terminal, high mass loss episodes. SN2002kg's eruption
was sub-luminous and it was quickly reognized as  a non-supernova. It 
was  identified with  the ``irregular blue variable'' V37 in NGC2403 \citep{ST1968}, an LBV/S Dor variable 
in its maximum light stage \citep{Weis05,VanDyk05a}. \citet{VanDyk06} and 
\citet{Maund06} have discussed its spectra and photometry during its 2002 - 2004 maximum. 

\begin{figure}
\figurenum{1}
\epsscale{0.7}
\plotone{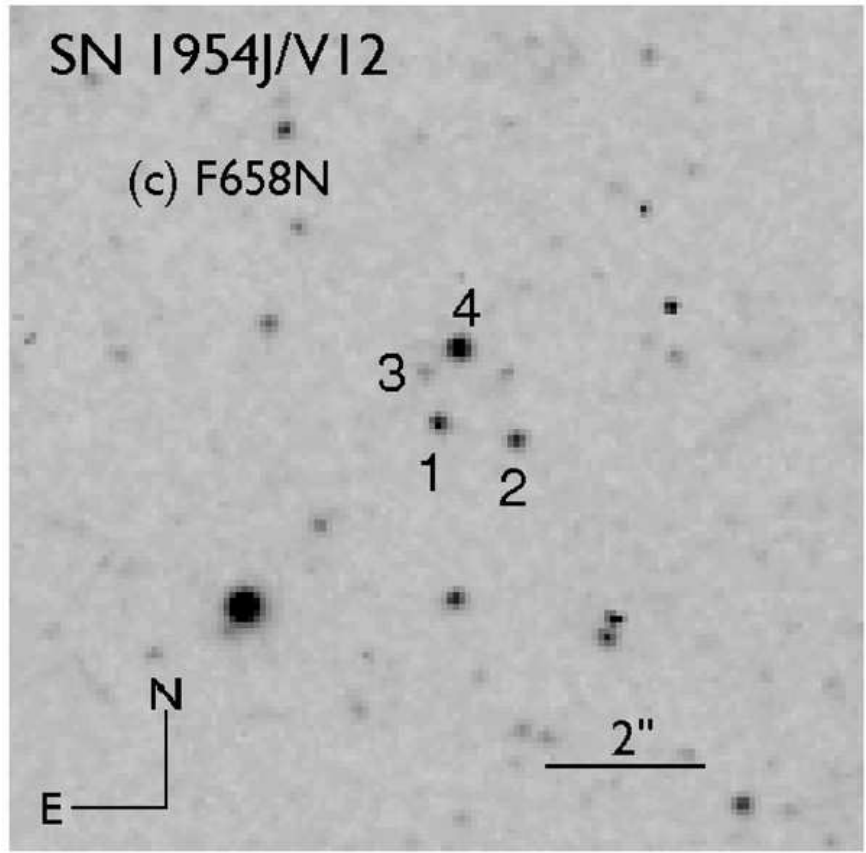}
\caption{The environment of V12/SN1954j. The H$\alpha$ {\it HST} image 
   from \citet{VanDyk05b} showing the four stars.}
\label{fig:environ}   
\end{figure}

SN1954J is famous. Based on its peculiar light curve \citep{ST1968} and its resemblance to $\eta$ 
Car and SN1961V, it was also considered another possible  example of Zwicky's  
Type V supernovae
\citep{Kowal}, but it is also known as V12 in  NGC 2403 \citep{ST1968}. Like V37, it  was also described as an irregular blue variable. \citet{HD94} and \citet{RMH99} included it as one of four examples of a giant eruption or $\eta$ Car-like variable.  HST imaging  \citep{VanDyk05b}  
 resolved V12 into four stars, one of which is very bright in H$\alpha$ \footnote{See Figure 8 in \citet{VanDyk05b}. The H$\alpha$ image is reproduced here in Figure \ref{fig:environ}.}.   
\citet{VanDyk05b} also obtained an echellette spectrum with the Keck-II telescope, albeit 
with low S/N, that showed an H$\alpha$ emission profile with 
broad wings similar to that of $\eta$ Car. They thus argued convincingly that this H$\alpha$ bright object, star 4, is most likely the survivor.  

Our new observations are described in the next section. In \S \ref{sec:v37} we 
discuss V37's  eruption and
post-eruption spectra, its luminosity, and  its  light curve. The analysis of V12's current spectrum and photometry and its light curve is less straightforward. In addition to the H$\alpha$ emission with an electron scattering profile, the red spectrum shows strong Ca II absorption lines typical of a G-type supergiant. V12's most puzzling characteristic is the decline in the luminosity of the apparent survivor compared with the pre-eruption object. We describe V12's pre-  and post- eruption characteristics, the evidence for circumstellar dust and the probable nature of the progenitor in \S {4}. In the final section we summarize our conclusions for the progenitors and surviving stars in 
each case.

\section{The Observations}  %%% ===-=== %%%  
\label{sec:obs}  

\subsection{Spectroscopy}   %%% ===-=== %%%
\label{subsec:spectr}  

Post-eruption spectra of V37 were observed   on  10 October 2012 and on 04 and 05 December 2012 with the Hectospec Multi-Object 
Spectrograph (MOS) \citep{Fab98,Fab05} on the 6.5-m MMT on Mt. Hopkins, as part of a larger program 
on massive stars in NGC 2403.  The Hectospec\footnote{http://www.cfa.harvard.edu/mmti/hectospec.html} has a 1$\arcdeg$ FOV and uses 300 fibers each with a core 
diameter of 250$\mu$m subtending 1$\farcs$5 on the sky. We used the 600 l/mm grating with the 
4800{\AA} tilt  yielding  $\approx$ 2500{\AA} coverage  with dispersion 0.54{\AA}/pixel and R $\sim$ 2000. 
 The same grating with a tilt of 6800{\AA} was used for the red 
spectra with $\approx$ 2500{\AA} coverage, 0.54{\AA}/pixel dispersion and R $\sim$ 3600.  
The total integration times were four hours  and three hours, in the blue and red 
respectively. 
The spectra  were reduced using an exportable version of the CfA/SAO SPECROAD package 
for  Hectospec data\footnote{External SPECROAD was developed by Juan Cabanela for use
on Linux or MacOS X systems outside of CfA. It is available online at:
http://iparrizar.mnstate.edu.}. The spectra were all bias subtracted, flatfielded and 
 wavelength calibrated.

A red spectrum  of V12  (star 4 in Figure \ref{fig:environ}) was  first observed with the 
MODS1 spectrograph on the Large Binocular Telescope (LBT) on 06 January 2014 with the 1\farcs0 wide  long slit and the 670 l/mm grating yielding wavelength coverage from  5000{\AA} to 1$\mu$m.  
V12 is very faint, V$\approx$23 mag  consequently, the purpose of this observation was to obtain a moderately
good S/N profile of its  H$\alpha$ line.
The total integration time was 45 minutes.
The spectra were bias subtracted and flatfielded using the MODS Spectral Reduction Package, and
were then extracted, and  wavelength and flux calibrated using the IRAF twodspec
and onedspec packages.

When the MODS2 spectrograph became available in 2016, we obtained a second set of spectra using both spectrographs, one on each mirror, to observe red spectra
with integration times of 90 minutes each, for a total of three hours 
yielding much
better signal to noise. The spectra were observed on 31 January 2017. 
The seeing was excellent at 0.6\arcsec consequently, the long slit 
was narrowed to 0.6\arcsec yielding improved spatial and spectral 
resolution, R $\sim$ 2300. The spectra were reduced as described above. 
They were wavelength calibrated using the night sky lines and flux calibrated 
with the standard star G191-B2B,  observed immediately before V12.

\subsection{Photometry}  %%% ===-=== %%%
\label{subsec:photom}

Previous images obtained with {\it HST} were used to derive magnitudes in our 
earlier papers on V37 \citep{VanDyk06} and V12 \citep{VanDyk05b}.  The new reduction package DOLPHOT \citep{Dolphin} has several advantages over DAOPHOT in these 
crowded fields. Consequently, we remeasured the images of V37 observed in 2004 
with ACS/WFC and ACS/HRC, and  the  ACS/WFC images of V12 from 2004 using DOLPHOT. 
We have also reduced the more recent  WFC3/UVIS and WFC3/IR images obtained in 2013 (GO-13477).  The revised and new {\it HST} photometry in Table 1 for V37  and Table 2
 for the  four resolved stars in V12's environment are flight system Vega magnitudes.  
DOLPHOT uses the VEGAMAG zero points from  \citet{Sirianni} for ACS/WFC and  from the WFC3 
web page\footnote{http://www.stsci.edu/hst/wfc3/phot{\_}zp{\_}lbn}, 
for infinite aperture. Note that the new DOLPHOT photometry yields fainter 
magnitudes in all case than the previously published data. 
The formal uncertainties reported by DOLPHOT for both WFC3 and ACS/WFC are quite good, 
0.01 -- 0.02 mag. The flight system magnitudes were transformed to the Johnson-Cousins system (BVI) using the 
conversion in \citet{Sirianni}.These should be considered somewhat uncertain for stars with stellar winds.  There is no current conversion of the WFC3 
magnitudes to the
Johnson-Cousins  system, so BVI magnitudes are only available for 2004. 

{\it Spitzer}/IRAC and MIPS observations of NGC 2403 were obtained as part of
several different programs from 2004 to 2008. Here we combine the different 
data sets to derive improved fluxes for V12 and V37.
We used the MOPEX (Makovoz \& Khan 2005; Makovoz \& Marleau 2005) package provided by the Spitzer Science Center to mosaic the individual artifact-corrected Basic Calibrated Data (cBCDs) and APEX Multi-Frame within MOPEX to perform point response function fitting photometry of point sources in NGC 2403.  The detection threshold in MOPEX/APEX was set to 3$\sigma$ in all bands. APEX detects a source near the SN 1954J position in IRAC channels 1 and 2 and at MIPS 24 microns, but not in IRAC channels 3 and 4. Similarly, no point source is detected at or near the SN 2002kg position in any of the bands. We establish the upper limits to detection based on the fluxes of detected stars at or near the detection limit in the immediate environments of both SN 1954J and SN 2002kg. The 24 micron region around V12 is complex with a variable background.   We therefore 
also treat the measured flux  as an upper limit.  The array-location-dependent correction, the pixel phase correction, and color correction were all applied to the IRAC detections.
The new fluxes are given in Table 3. Our measurements while not identical, are consistent with those published by \citet{CSK12}.  

%%% ===-=== %%%  
\section{V37(SN2002kg), an LBV/S Dor Variable}  %%% ===-=== %%%  
\label{sec:v37}  

V37/SN2002kg was neither a supernova nor a giant eruption like SN1954J. 
It was discovered on 22 October, 2003 \citep{Schwarz03}, but first 
became visible a year earlier in images from  26 October 2002 obtained with the Katzman Automatic 
Imaging Telescope. It was initially classified as a Type IIn supernova \citep{Filip03}
because of its narrow hydrogen emission lines. 
SN2002kg was later identified with the irregular blue
variable V37 in NGC 2403 \citep{ST1968} and shown to be an LBV/S Dor variable in eruption \citep{Weis05,VanDyk05b}.

\subsection{The Eruption and Post-Eruption Spectra}  %%% ===-=== %%%  
\label{subsec:v37erupt}  

\begin{figure} 
\figurenum{2}
\epsscale{1.0}
\plotone{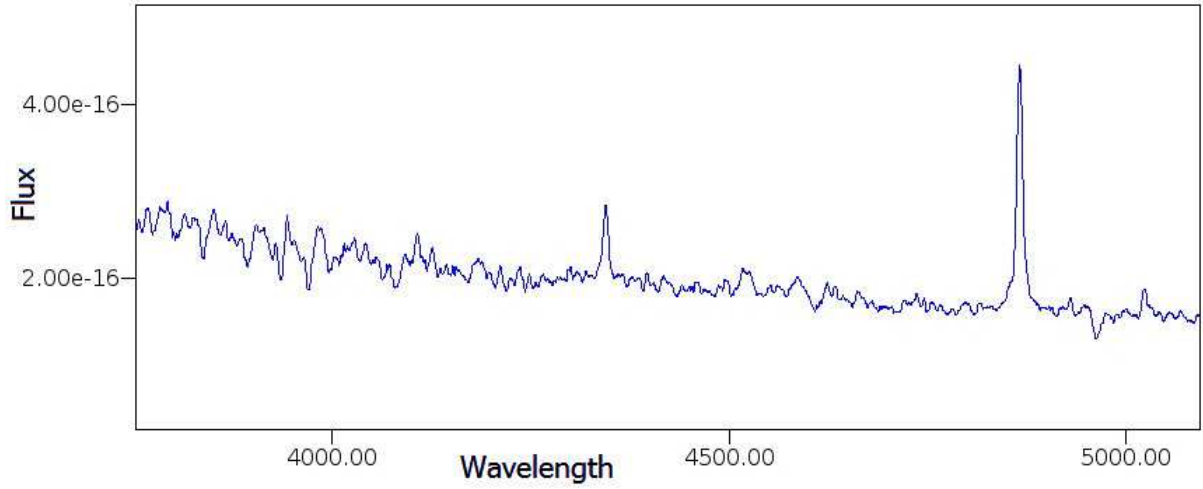}
\caption{The flux calibrated (ergs cm$^{-2}$ s$^{-1}$ A$^{-1}$) 2003 
   spectrum of SN2002kg (V37).}
\label{fig:v37spectr2003}
\end{figure}

Spectra observed during maximum light, 2003 - 2004, showed absorption lines 
characteristic of an early A-type supergiant \citep{VanDyk06,Maund06}. Major LBV 
outbursts usually develop strong F-type absorption-line  spectra. 
This difference only means
that V37's dense wind  did not get as cool as recent examples such as 
 Var C in M33 \citep{RMH14a} and R71 in the LMC \citep{Mehner}.
Although, during its first recorded S Dor eruption in the 1970's, R71 also developed an
A-type absorption line spectrum \citep{Wolf81}.
\citet{VanDyk06} published four spectra and \citet{Maund06} published two spectra covering the two year period from 
January 2003 to December 2004. The spectrum  from  the MMT blue spectrograph observed on
20 November 2003 is shown in Figure \ref{fig:v37spectr2003}. Strong absorption lines of hydrogen, H$\epsilon$, $\lambda$3835{\AA}, and $\lambda$3889{\AA}, and the Ca II K line  are conspicuous as are the hydrogen emission lines of H$\alpha$, H$\beta$, H$\gamma$ and H$\delta$. Emission lines of 
Fe II and [Fe II] are also present. Although no P Cygni profiles are apparent, 
electron scattering wings common in strong stellar winds are observed on the 
hydrogen emission profiles.  

\begin{figure}
\figurenum{3}
\epsscale{1.0}
\plotone{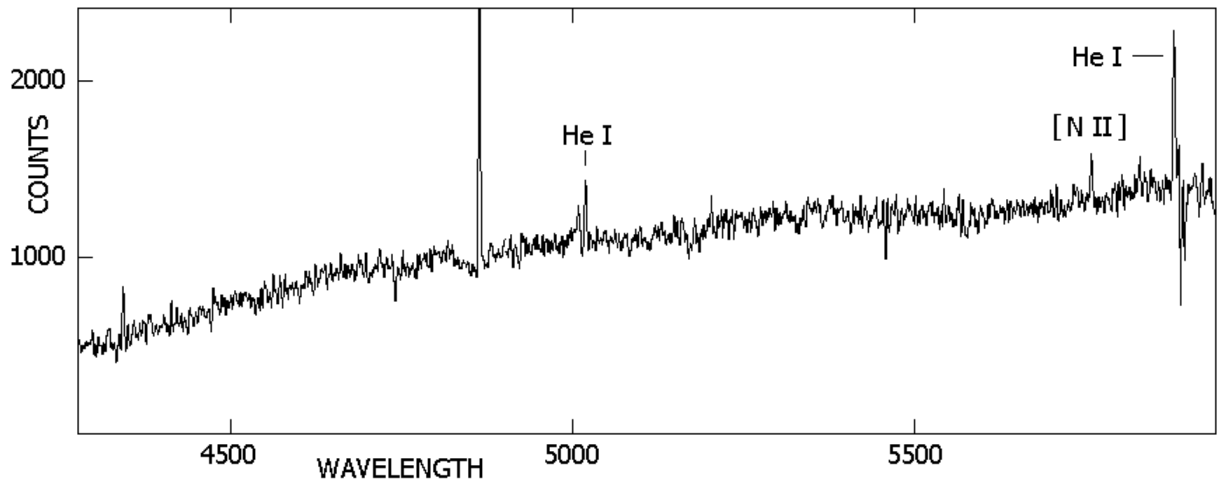}
\caption{The blue spectrum of V37 from 2013. The He I $\lambda$5876 lines 
    and the [N II] line at $\lambda$5755 are marked.}
\label{fig:v37bluesp2013} 
\end{figure}

\begin{figure}
\figurenum{4}
\epsscale{1.0}
\plotone{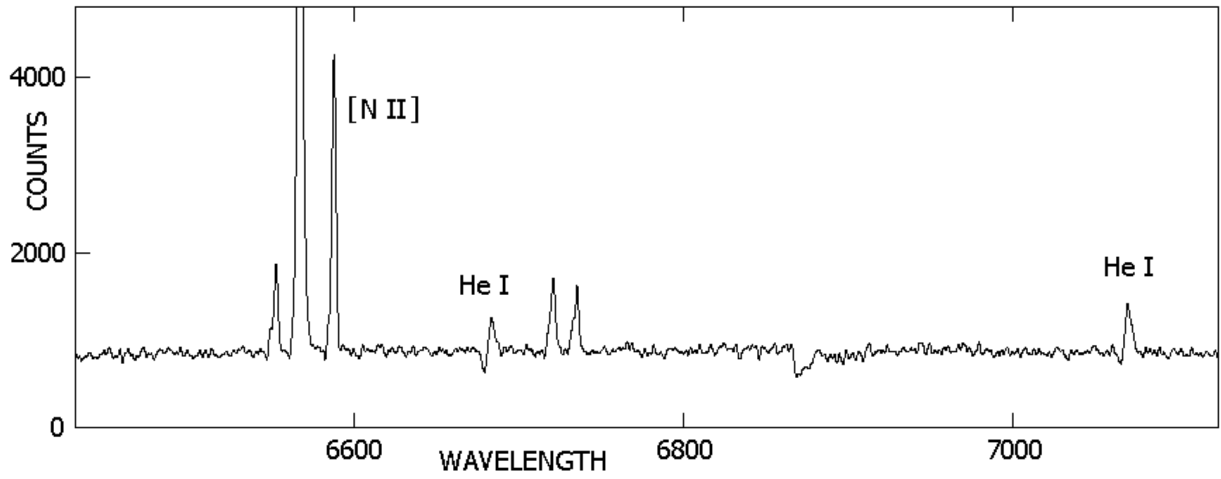}
\caption{The red spectrum of V37 from 2013. The [N II] line at 
     $\lambda$6584 and two He I lines with P Cygni profiles are marked.}
\label{fig:v37redsp2013}  
\end{figure}

The blue and red spectra of V37 from  2013 are shown in Figures \ref{fig:v37bluesp2013} and \ref{fig:v37redsp2013}. The star 
is quite faint and the spectra have low S/N. In addition to the hydrogen emission lines  the spectra show prominent P Cygni profiles in the   He I emission lines at $\lambda\lambda$ 4471,5015,5876,6678, and 7065{\AA}. The   outflow or wind 
speed measured from  the absorption minima in the P Cygni profiles in this low S/N 
spectrum is $-$187 $\pm$ 10 km s$^{-1}$. This  velocity is consistent with the low wind speeds 
measured similarly for the LBVs in M31 and M33 by \citet{RMH14b} who reported that the winds of 
LBVs even in their quiescent state are slower than for normal hot supergiants. 

The more recent post-eruption spectra also show the nebular emission lines of [N II], [S II] and a relatively strong
[N II] $\lambda$5755 line.  The ratio of the [N II] lines 
$\lambda$5755/$\lambda$6584 indicates electron densities on the order of  
10$^{6}$ cm$^{-3}$, greater than for an H II region. This suggests  the presence 
of circumstellar nebulosity which very likely  formed during its 2002 - 2004 eruption. 

\subsection{The Luminosity Before  and During the Eruption}  %%% ===-=== %%%
\label{subsec:v37lum}  

Pre-eruption photometry on the 
UBVRI photometric system  \citep{VanDyk06} is available from  images  obtained with the Nordic Optical Telescope on 13 October 1997 \citep{Larsen99} and on the {\it g$^{'}$r$^{'}$i$^{'}$} system from the Isaac Newton Telescope on  20 February 2001 \citep{Maund06,Maund08}. The pre-eruption 
magnitudes are summarized in Table 1. The interstellar reddening and visual extinction were independently derived from the two-color diagram \citep{VanDyk06} and from stars within 2$\arcsec$ \citep{Maund06}. Both approaches give approximately the same results with E$_{B-V}$ of 0.19 mag and 0.17 mag, respectively (Table 1). Since the colors of luminous hot stars may be affected by 
emission  lines, we adopt the reddening from the nearby stars (A$_{V}$ = 0.54 mag), and using the \citet{Cardelli} extinction law we derive the intrinsic colors in Table 1. The 2001 B-V  color \citep{Maund08} is much redder than the earlier photometry. Since those data were obtained only a short time before the eruption was 
detected, it is likely that V37 was already in transit to the cooler, maximum light state, 
although we would have expected the  star to have begun to brighten in  V.  \citet{Maund06} commented that
the colors in their pre-eruption images were poorly constrained by the g$^{'}$ and i$^{'}$ band images. For that reason we adopt the 1997 photometry and colors from the NOT for the pre-eruption
star. 

V37's pre-eruption intrinsic colors correspond to those of an early-type hot supergiant with
an approximate spectral type of O9-B0. At a
distance modulus of 27.5 mag \citep{Freedman} for NGC 2403, its M$_{V}$ is $-$7.5 mag corrected for interstellar extinction. Adopting the 
corresponding bolometric correction \citep{Flower96,Martins}, the  M$_{Bol}$ of the pre-eruption 
star would be  
$-$10.4 mag.  During its outburst, near maximum light, M$_{V}$ was  $-$9.8 mag from  the 26 March 2003
photometry \citep{VanDyk06} with the same A$_{V}$. The unreddened colors  
also correspond to a late B or early A-type supergiant consistent with the appearance of its spectrum during the eruption. With its corresponding bolometric correction, its luminosity was then M$_{Bol}$ $-$10.4 mag near maximum.  
Based on stellar models with and without rotation \citep{Ekstrom}, V37's initial mass would have been 60 - 80 M$_{\odot}$. 

Thus V37's outburst was 
like that of other LBV/S Dor variables, an apparent 
transit on the HRD, at nearly constant luminosity, due to increased mass loss and the 
formation of a cooler, dense wind. Also like other LBVs \citep{RMH14b,Kraus,Oksala}, 
V37 apparently did not form  dust  associated with its eruption \citep{CSK12}, although it has apparently formed  a post-eruption circumstellar nebula.

\subsection{The Light Curve}  %%% ===-=== %%%  
\label{subsec:v37lc}   

\begin{figure}
\figurenum{5}
\epsscale{1.0}
\plotone{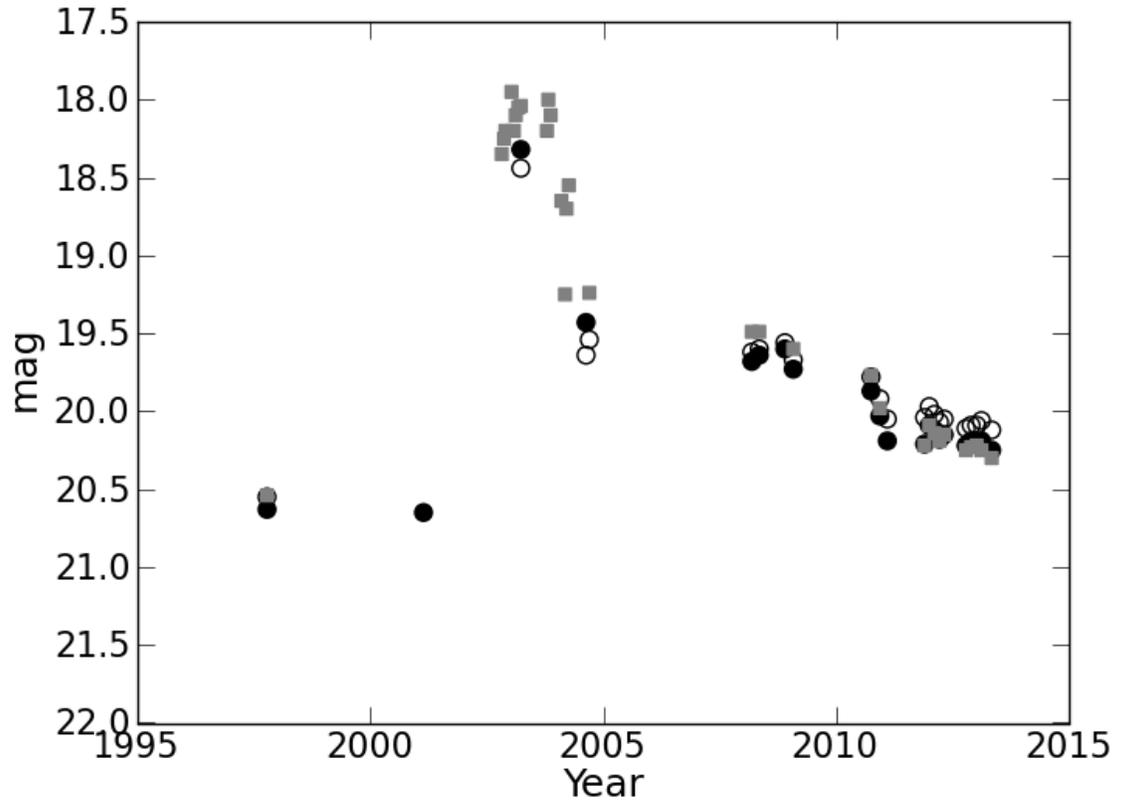}
\caption{Light  curve for V37, 1997--2013.  Filled circles are the v-band 
    magnitudes, open circles are b-band and grey squares are r-band. }
\label{fig:v37lcurve}  
\end{figure}

V37's historical light curve can be seen in \citet{ST1968}, and in \citet{Weis05}. 
We show V37's light curve from its current eruption in Figure \ref{fig:v37lcurve} with the V- and B-band magnitudes from Table 1, the 
revised B magnitudes using DOLPHOT with the  HST/ACS images from 2004, plus the R band magnitudes from \citet{VanDyk06}  to illustrate 
its  maximum light. In addition, \citet{CSK12} have published multi-color 
photometry from 2008 - 2011 from the LBT/LBC. The LBT/LBC photometry 
in poor seeing may have included light from  a nearby star about 1 arcsec away. See their Figure 20. We therefore measured the brightness of this star in the ACS images and conclude that at $\approx$ 22 mag the fainter star does not contribute
significantly to the magnitudes for V37. Their BVR  band magnitudes are 
also included on Figure \ref{fig:v37lcurve} plus addtional LBT/LBC photometry from 2011 to 2013 (Gerke 2013). The post eruption photometry shows that V37 was relatively constant from about 2004 to 2008 followed by  a slow decrease until about 2010  
when it began a more rapid decline.

\section{V12 (SN1954J) --  a Giant Eruption}  %%% ===-=== %%%  
\label{sec:v12}  

\begin{figure}
\figurenum{6}
\epsscale{1.0}
\plotone{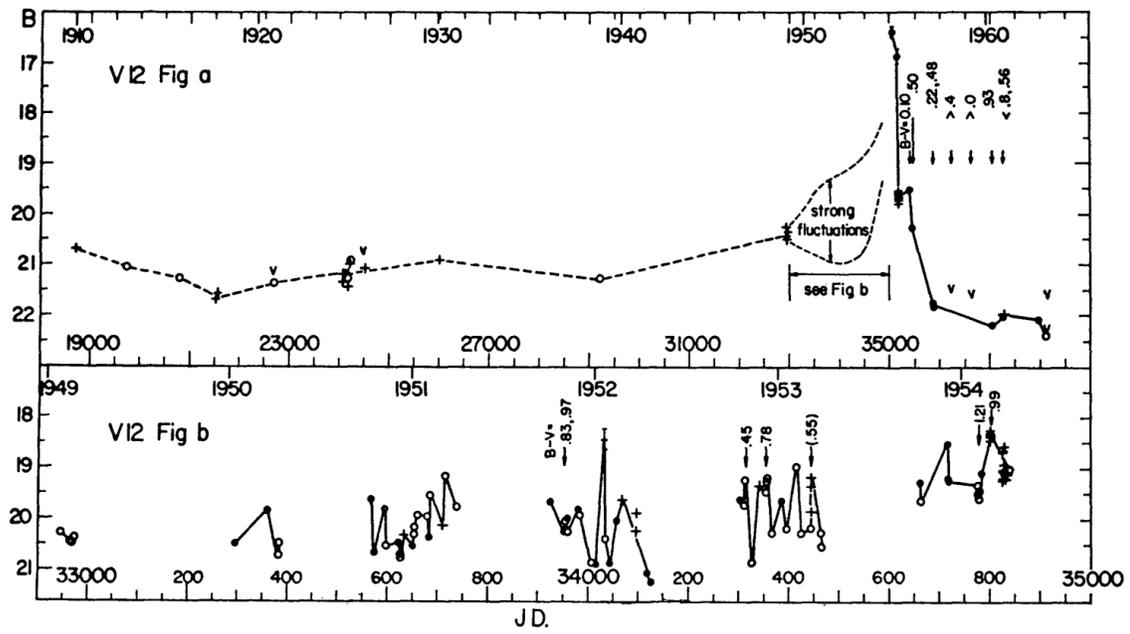}
\caption{Historical light curve for V12 (SN1954J) from \citet{ST1968}.}
\label{fig:v12lc}
\end{figure}

V12 increased at least five magnitudes in apparent brightness during its eruption. Its historical light curve showing its 1954 eruption \citep{ST1968} is reproduced in Figure \ref{fig:v12lc}.  Its  pre-eruption light curve is remarkable for the  rapid variations in its apparent magnitude in the years 1950 -1954 prior  to its giant 
eruption, sometimes  a half magnitude or more in only a few days. These erratic fluctuations are most likely due to a short term  surface instability  distinct from  the  LBV/S Dor long-term  eruptions which have different characteristics. 
 $\eta$ Car also displayed rapid oscillations prior to its ``great eruption'' in 1843. Similar variability  was also reported in the pre-outburst light curve of the SN impostor UGC 2773-OT \citep{Smith2010}, and multiple outbursts have been observed in the impostor SN 2000ch \citep{Wagner,Pastorello} as well 
 as SN 2009ip before its presumed terminal explosion \citep{Pastorello13,Mauerhan,Margutti}.  
 These geyser-like lesser outbursts may be clues to the cause of  
 the giant eruption, although the origin of  these instabilities is likewise not understood.       

These short term variations in V12 were superposed on a rather slow increase in
brightness from about 20.5 mag in late 1949 to 19.1 mag in March 1954.  
There is then a seven month gap in its photometric record because of its 
proximity to the Sun at the time. 
When next observed in November 1954, it was at apparent B magnitude 16.46, 
after which it rapidly declined with 
an apparent  short plateau or shoulder on its light curve that lasted another seven 
months \footnote{The apparent plateau corresponds to a second  gap 
in the observatons when NGC 2430 was not observed due to its position 
on the sky.  On 29-30 March 1955 and on 21-22 Oct 1955, V12 was measured at 19.6 and 19.5 mag, respectively, hence the conclusion that it had not varied significantly during that time. The magnitudes from  \citet{ST1968} were measured from photographic plates and there is always some question about the associated errors or uncertainties. \citet{ST1968} used a primary photoelectric sequence on the photographic plates. Based on their discussion, the error for the sequence stars brighter than 19.5 mag is $\pm$ 0.02 mag. For stars fainter than 21 in B it is $\approx$ $\pm$ 0.1 mag. Another way to estimate the likely error specifically for V12, is to compare measurements separated by only a few days
when it was not varying rapidly. We find a scatter typically $\pm$ 0.05 -- 0.10 mag in B when it was a 21st mag star,  and $\pm$ 0.10 mag in V from  their Tables A3 and A4, respectively which supports internal consistency.} 
A short shoulder on the declining light curve is also observed in the light curves of
other giant eruptions including P Cyg, SN1961V, and probably $\eta$ Car \citep{RMH99}. 
\citet{Mauerhanb} have suggested that these plateaus, which look like a brief 
stall in the
declining light curve, are due to interaction of the expelled mass with previously ejected
circumstellar material. If correct, this is evidence that V12 may have had previous high mass loss events. 

Unfortunately we do not know its luminosity at maximum.
Based on the observed magnitudes, V12 reached an absolute B magnitude  of at least M$_{B}$ $= -$11.0.  Correcting for possible 
insterstellar extinction of A$_{V}$ $\approx$ 0.9 mag (1.2 mag A$_{B}$), based on the discussion in \S {4.2}, M$_{Bmax}$ would be at least -12.2 mag. Although no color was recorded
at maximum light, its color reported several times during its rise 
(1951 -- 1954) was $B-V \simeq 1.0$, consistent with the production of a 
cool, dense wind with a negligible bolometric correction.  
Adopting this color, and correcting for 
extinction, then M$_{V}$ $\approxeq$ M$_{Bol}$ at maximum was $-$12.9 mag, or greater; comparable to several other SN impostors with reported maximum luminosities on the order of $-$13 to  $-$14 mag.

\subsection{The Post-Eruption Spectrum -- H$\alpha$ and Absorption Lines}
\label{subsec:v12spectr}   %%% ===-=== %%%   

\begin{figure}
\figurenum{7}
\epsscale{1.2}
\plotone{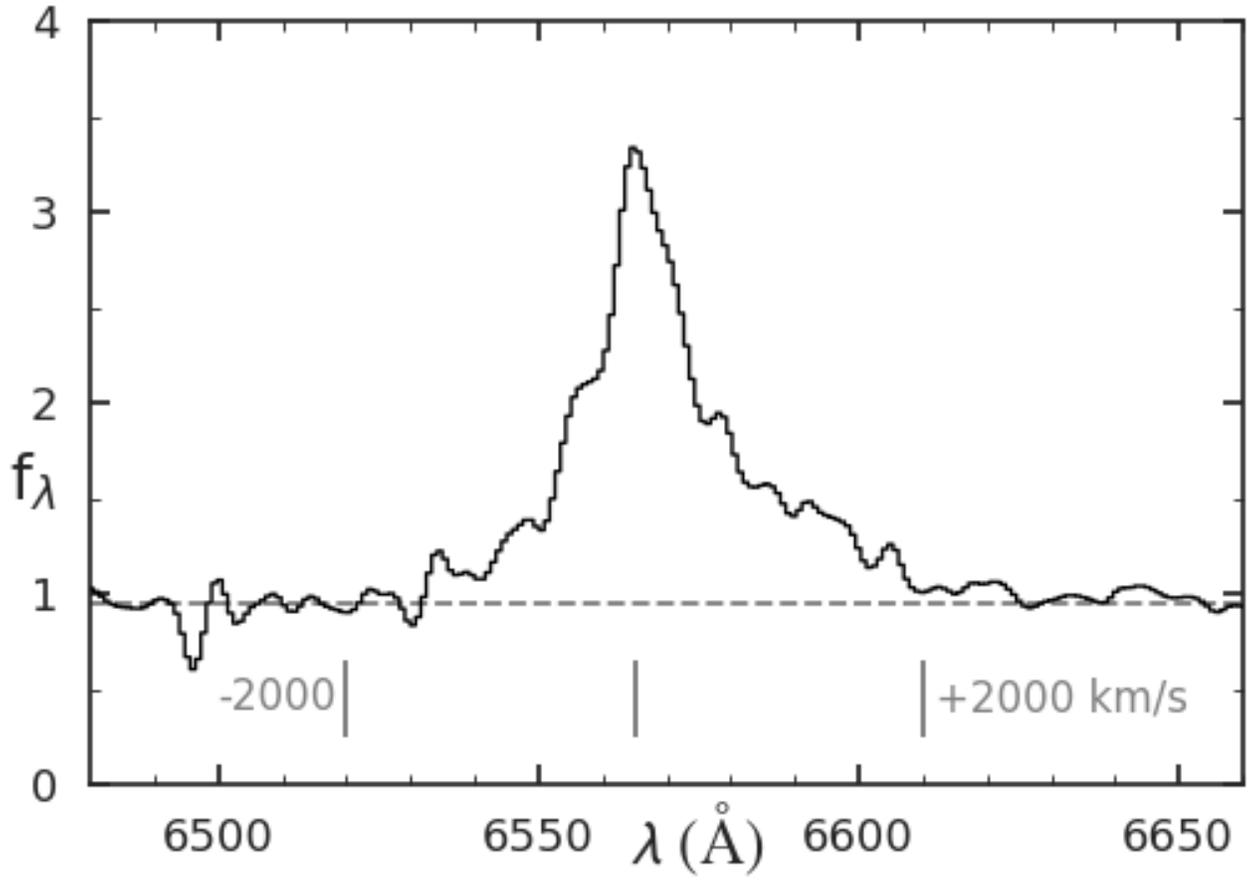}  
\caption{The flux calibrated H$\alpha$ profile of V12 in 2017 January. The units for f$_{\lambda}$ are  10$^{-17}$ ergs cm$^{-2}$ s$^{-1}$ A$^{-1}$.}
\label{fig:halpha}    	
\end{figure}

The H$\alpha$ emission profile from our 2017 January LBT/MODS1+MODS2  
spectrum  of star 4 is shown in Figure \ref{fig:halpha}.
This is the line profile of a stellar wind, not a supernova remnant. 
It has broad asymmetric wings characteristic of Thomson scattering, unrelated 
to bulk motion of the gas, and there is no obvious  P Cygni absorption feature.  
The line profile's  total equivalent width is 59{\AA} and its net flux corrected 
for interstellar extinction (\S {4.2}) is 1.1 $\times 10^{-15}$ ergs cm$^{-2}$ s$^{-1}$ (1.1 $\times 10^{-18}$ W m$^{-2}$),  about  330 $L_{\odot}$ at the distance 
of NGC 2403. There is no significant change in the flux from the 2014 spectrum. Its FWHM line  width  of  13.5{\AA}  
 corresponds to a velocity of 620 km s$^{-1}$ similar to  $\eta$ Car's value of 
 600 km s$^{-1}$. This may be an overestimate because the  
profile is already  asymmetric at the half maximum level (2.17 $\times 10^{-17}$ ergs cm$^{-2}$ s$^{-1}$). 
 Before the profile begins to noticeably 
 broaden, the width is about half that, equivalent to 270 km s$^{-1}$, measured at  2.84 $\times 10^{-17}$ ergs cm$^{-2}$ s$^{-1}$.
The asymmetric wings extend to $-1530$ km s$^{-1}$ and 
$+2000$ km s$^{-1}$.  The Heliocentric velocity of the peak of the H$\alpha$ 
line is  133 km s$^{-1}$.

The  red spectrum also reveals several absorption lines shown in Figure \ref{fig:caii}. The 
near-infrared Ca II triplet is readily identified. The lack of emission and 
P Cygni profiles in the Ca II lines, often seen in the spectra of cool stars with high mass loss and circumstellar ejecta \citep{RMH13,Gordon}, is somewhat surprising. Several other 
absorption lines of Fe~I, Ti~I the Mg~I triplet at $\lambda$8806{\AA} and the Si I 
doublet at $\approx$ 8751{\AA} are also present. These neutral metallic lines 
imply that the photosphere or very dense wind is relatively cool, corresponding 
to a star later than spectral type F.  The lack of a strong O I $\lambda$ 7774 line also requires a later spectral type. We use the Ca II spectral library \citep{Cenarro}\footnote{http://pendientedemigracion.ucm.es/info/Astrof/ellipt/CATRIPLET.html} to estimate a mid to late  G spectral type based primarily on the strength of the 
Ca II lines and the neutral lines. The Fe I and Ti I lines  
all show increasing strength with luminosity. The line ratio 
Fe I $\lambda$8688 to Ti I $\lambda$8675 is a known luminosity 
indicator \citep{Keenan}. Our measured ratio, 1.5 -- 1.9, for this line pair 
in V12 compared with ratios of 1.3 to 2.0  measured in several G-type supergiants
in the spectral library  suggests
a relatively high luminosity  for the G-type star. We also note that 
the equivalent width of the luminosity sensitive O I $\lambda$ 7774 line is the same as that measured in a G5 supergiant in M31 \citep{Gordon}. Based on its 
spectral characteristics we would expect an apparent temperature of $\approx$
 5000 K.  The average Heliocentric velocity of the Ca II lines is 138 $\pm$ 1 km s$^{-1}$ compared with 133 km s$^{-1}$ for the 
H$\alpha$ peak.The systemic velocity of NGC 2403 is 133 km s$^{-1}$. V12 is located very close to the minor axis of the galaxy and only $\approx$ 1.5 kpc from the center so it is not surprising that the measured velocities are close to the systemic.

\subsection{Interstellar Extinction and the Spectral Energy Distribution} 
\label{subsec:v12extinct}   %%% ===-=== %%% 

{\it HST} imaging resolves V12 into 
four stars (Figure \ref{fig:environ}). The strong H$\alpha$ source, star 4, is the most likely candidate for the surviving star. 
 The {\it HST} based VEGAMAGS in Table 2, separated by nine years,  also show that none of 
the four stars has varied significantly in the past decade. Assuming that these stars are also members of 
 NGC 2403, they make up the immediate environment of V12 and can be used to
estimate the internal interstellar extinction. \citet{VanDyk05b} discussed the two-color
B-V vs V-I diagram  using their earlier DAOPHOT photometry, and concluded that star 3 is
most likely a reddened hot star and the two redder stars are  red supergiants.   

We use the revised magnitudes
from  DOLPHOT  for the 2004 ACS data for the two-color diagram (F475W-F606W vs F606W-F814W), in Figure \ref{fig:2color}. It doesn't look much different from their earlier figure except that the errors are much smaller with DOLPHOT. The smooth curve is the intrinsic color locus for supergiants based on the Castelli-Kurucz models \citep{Castelli}, convolved with the ACS/WFC filter functions using STSDAS/SYNPHOT. The dotted line from the tip (hottest) of the supergiant locus is the reddening vector 
for A$_{V}$ = 1.0 mag using the \citet{Cardelli} interstellar extinction curve.  
Star 4 lies blueward of the intrinsic color curve. Indeed its colors are anomalous; it is too blue for its apparent G spectral type.  Star 3 lies close to the intrinsic color  line. Its colors suggest a hot star with A$_{V}$  $\approx$ 0.9 mag., or alternatively an  intermediate temperature star (~B8 I-A0 I) with only the  Galactic foreground extinction  of 0.11 mag \citep{SF2011}, and therefore virtually no internal interstellar extinction from NGC 2403. But, the colors of the two red supergiants also indicate some additional internal extinction; they would lie on the intrinsic color with an A$_{V}$ of $\approx$ 0.8 mag.  
Given V12's location in an association of young stars in a spiral arm, some internal reddening is expected.  For example, V37 discussed previously has an A$_{V}$ of 0.54 mag. Our results suggest an A$_{V}$ of 0.8 - 0.9 mag for the V12 environment.  

Using the BVI magnitudes in Table 2 with A$_{V}$ of 0.9,  star 4
 has  a current  M$_{v}$ of $-$5.0 mag  with an intrinsic $B-V$ of $\approx$  0.76 mag.,  suggestive of an intermediate temperature star with a corresponding  M$_{Bol}$ of $-$5.1 mag, only $\approx$ 8 $\times 10^{3} L_{\odot}$. The derived intrinsic color suggests a mid F-type star which is not supported by
   the absorption line spectrum. The expected intrinsic $B-V$ color for a mid to
      late type G supergiant is  0.9 -- 1.0  mag. The derived luminosity is also 
      discrepant with the luminosity indicators in its spectrum.
          This relatively low  luminosity would place star 4
	        well below the AGB limit on the HRD.

Star 4's current spectral energy distribution (SED) is shown  in Figure 10 
togther 
with the SEDs for the two red stars combined and for star 3 
for comparison.  
We use the fluxes for the flight system magnitudes in Table 2\footnote{We use 
the Vega flux calibration for the
ACS/WFC and WFC3 filters at http://www.stsci.edu/hst/acs/analysis/zeropoints/{\#
}vega and at http://www.stsci.edu/hst/wfc3/phot{\_}zp{\_}lbn with the pivot 
wavelengths of the filters.}.  The observed magnitudes are corrected for the adopted interstellar extinction using the \citet{Cardelli} extinction curve. 
The contribution  
from the H$\alpha$ line has been removed from the F606W  flux for star 4.
The  {\it Spitzer}/IRAC and MIPS measurements from Table 3 will include the light from all four stars and are therefore shown on all three SEDs. They reveal a 
near-infrared source at 3.6$\mu$m and 4.5$\mu$m, upper 
limits at 5.8$\mu$m and 8.0$\mu $m, and an upper limit on a  mid-infrared 
source at 24$\mu$m 
from MIPS. 

Planck curve  fits are shown  for all three SEDs.  
For the purposes of this paper, the  Planck curve 
 adequately approximates the continuum  at blue to far-red 
wavelengths -- i.e., the  temperatures are meaningful even though they differ a little 
      from the formal effective temperature $T_\mathrm{eff}$.\footnote{Elaborate    
    photosphere models would require more parameters than our data 
      can support, and, besides, are beyond the scope of this paper. For $\lambda \lesssim 0.37 \; \mu$m, the Balmer discontinuity may 
        indirectly cause our blue-to-red Planck-fit temperatures to 
	  exceed the the true $T_\mathrm{eff}$, but this effect is not large 
	    for supergiants, and the Paschen jump is smaller.      
	      At $\lambda \gtrsim 0.8 \; \mu$m, free-free emission from a stellar 
	        envelope or wind may supplement the continuum.}   

The two red stars have nearly identical magnitudes and colors and are clearly the 
dominant contributors to the near-IR flux as illustrated by their combined  SED
(Figure \ref{fig:sed1and2}). 
Star 3 is a hot supergiant. The Planck curve suggests a temperature of 15000 to 20000 K with a small free-free excess (Figure \ref{fig:sed3}). The best Planck 
curve fit to star 4 (Figure \ref{fig:sed3}) however, indicates a much warmer 
temperature, $\sim$ 6600 K, corresponding to a mid F-type supergiant, than 
indicated by the absorption lines. A 5000 K 
blackbody is shown  for comparison as are the magnitudes for the pre-eruption star.
The likelihood that star 4 is actually two stars is discussed more thoroughly 
in \S {4.4} .

 \subsection{The Nature  of the Progenitor}
\label{subsec:v12nature}  %%% ===-=== %%%  

V12's historical  light curve (Figure \ref{fig:v12lc}) shows that the progenitor was significantly brighter than the apparent survivor is today.
 Prior to its erratic rise to maximum,
 V12 had an average pre-eruption (1910 - 1938) photographic or B magnitude of 21.2 mag
 from the  F(air) and G(ood) quality magnitudes from  Table A3 in \citet{ST1968}.
 These photographic magnitudes refer to the integrated light of the four stars resolved by
 {\it HST}. Today, the combined apparent B magnitude of the four stars in Table 2 is 22.55 mag, 1.3 mag fainter than in its pre-eruption state. Assuming that the other three stars
 had the same blue magnitudes they have today, then star 4 would have had to be a 
 21.5 magnitude star prior to the eruption, 2.5  to 3  magnitudes brighter than its current blue magnitude. Adopting A$_{V}$ of 0.9 mag (A$_{B}$ of 1.2 mag),
 its absolute blue magnitude then was $-$7.2. Assuming the $B-V$ $\cong$ 1.0 mag color observed frequently during the rise to maximum, 
 its M$_{V}$ would be $-$7.9 mag and with no bolometric correction, the progenitor would be an intermediate temperature supergiant with a luminosity of  $\approx$ 10$^{5} L_{\odot}$ and  an initial mass of $\approx$ 20 M $_{\odot}$. Of course if the star was hotter, it would be more luminous. 

Thus the likely survivor, star 4, is now  approximately 2.5 to 3 magnitudes 
fainter than its probable progenitor, and its luminosity today is at least 
10 times fainter.  This is a surprising result. As the star recovers from a giant 
eruption it is expected restore hydrostatic and thermal equilibrium. We 
 therefore expect the 
star to return to close to its pre-eruption luminosity. 
$\eta$ Car is the best example we have of a star 
recovering from a giant eruption, and the data on its pre-eruption state support
 very 
 little or no change in its total luminosity \citep{KD97}.

Some may argue that there is no problem; the SN1954J eruption  was terminal, i.e. a true supernova,  the star is gone, and  the object we now observe is a completely 
different star. However, 60 years later,  there is a no evidence for a SN remnant at the position of star 4. There is no X-ray source \citep{Binder} or non-thermal radio source \citep{Eck} at or near its position, and the H$\alpha$ profile is representative  of a stellar wind.

Another possibility is that star 4 is actually more than one star. 
Star 4 in the F658N image has a Moffat FWHM of 1.47 pixel, while both stars 1 
and 2, interestingly, have FWHM of 2.12 and 2.01, respectively. The PSF for star 4 is 
consistent with what would be considered stellar, or at least, unresolved.
The  corresponding region sampled with ACS/WFC is only  1.1 pc at the 
distance of N2403. Of course, this does not rule out the presence of another
star in the PSF. As already mentioned star 4's colors are too blue for the G spectral type. V12 could also be a binary or a multiple system, but 
that would not solve the problem  of the decrease in
apparent brightness and luminosity. Possible models for star 4 are discussed
below. 

\subsection{Conceptual Models  for Star 4}         
\label{subsec:models} 

It is surprisingly difficult to assemble a model that explains all the 
data on V12.  The H$\alpha$ emission line, for instance, causes trouble 
in the following way.    Its profile (Fig.\ \ref{fig:halpha}) 
has the classic shape produced by Thomson scattering in a mass outflow, 
with  asymmetric quasi-exponential wings extending to $-$1500 and $+$2000 
km s$^{-1}$  -- see, e.g., \citet{auervanb,kd95,dessart,HD2012}.  The 
long-wavelength wing appears consistent with average optical depth 
$\tau_\mathrm{sc} \gtrsim 1$ in the H$\alpha$ emission region.   
Since line profiles of this type characteristically occur in giant 
eruptions and other very dense outflows, this interpretation seems  
natural for V12 -- but the simplest models predict inadequate 
values of $\tau_\mathrm{sc}$.   Below we sketch the most evident 
possibilities for star 4.

\begin{enumerate}
  %%%  
\item   Suppose that this object is currently a supergiant with 
$L \gtrsim 10^4 \ L_\odot$, $T_\mathrm{eff} \approx 5000$ K, and 
$R \gtrsim 130 \ R_\odot$.  The temperature is indicated by the 
 \ion{Ca}{2} triplet and other absorption features 
 ({\S}\ref{subsec:v12spectr}). 
 But the SED shown in Fig.\ \ref{fig:sed4} requires $T \approx 6600$ K,  
 a temperature difference that is spectroscopically unacceptable. 
Moreover, a single-supergiant model  
 fails to explain the H$\alpha$ profile.  In order to produce  
 the observed H$\alpha$ luminosity in an envelope or wind around 
 such a large star, the ionized density just outside radius $R$ 
 must be of the order of $n_\mathrm{e}  \sim  10^{10}$ cm$^{-3}$.  
The resulting optical thickness for Thomson scattering 
is only ${\tau}_\mathrm{sc} \sim 0.1$, which is too 
 small to create the H$\alpha$ line wings.  This difficulty is   
 basically a consequence of the star's large radius.  
 The discrepancy 
 worsens if we allow for inhomogeneous densities, and circumstellar  
 extinction ({\S} \ref{subsec:qmodels}) would not help.  Hence, in 
 this model the H$\alpha$ line shape must be caused by an unusual 
 distribution of bulk outflow velocities, with a maximum that is 
 much larger than the escape velocity (roughly 2000 km s$^{-1}$  
  vs.\ 200 km s$^{-1}$ respectively).  In summary, this object 
 appears more complicated than a single G supergiant.  
  %%% 
  \item  The most obvious possibility is to add another, hotter star.  
  There is nothing implausible about either a binary companion, 
  or an unrelated object within about 1 pc.  As already emphasized, 
  star 4's SED cannot be fit by a single  5000 K star. 
  The observed spectrum sets limits on the hypothetical second  
  object.   The lack of He I emission 
  ($\lambda$5876 and $\lambda$6678) in a star with a strong  
  stellar wind suggests that it is not likely to be much hotter than 
  20000 K, while a 7000 --- 10,000 K object would contribute 
  features not observed in the red spectrum.  A combination of two 
  stars can account for both the blue-to-red SED and the 
  \ion{Ca}{2} triplet near 8500 {\AA}.  Fig.\ \ref{fig:2comp} 
  shows an example with temperatures of 5000 K and 20000 K;  
  note that the cooler object must dominate around 0.85 $\mu$m so 
  \ion{Ca}{2} can be prominent.  
  The giant eruption survivor is presumably the hotter object,  
  because it can have an unusually dense wind to account for 
  the brightness  and profile of the H$\alpha$ emission.  Since 
  it is much smaller than the G supergiant, its wind can have 
  $\tau_\mathrm{sc} \sim 1$ while producing the right amount of 
  H$\alpha$.    
  Incidentally, there is no need to postulate a close 
  {\it interacting\/} binary.  More details are outlined in 
  {\S} \ref{subsec:qmodels} below.
  %%%%
  \item  Other scenarios are more complicated. 
  For instance,  the H$\alpha$ might originate in ejecta from the 
  1954 eruption, now located at $r \sim$ 4000 to 8000 AU assuming expansion 
  velocities of 300--600 km s$^{-1}$ based on the width of the 
  H$\alpha$ line.  Such models can be devised for ejecta masses   
  in the range 0.1 to 10 $M_\odot$, with suitable densities and  
  volume filling factors.  Thomson scattering is then negligible,  
  so the H$\alpha$ profile must represent line-of-sight bulk expansion 
  velocities.  This is unsatisfying, because it does 
  not account for the \ion{Ca}{2} vs.\ SED mismatch noted above,  
  and also because the ejecta velocity distribution 
  fortuitously mimics a  Thomson-scattered profile.  
  The  observed asymmetry in the profile (red side brighter) matches 
  Thomson scattering but is opposite to what one normally expects 
  for a Doppler profile in expanding ejecta.   
  %%% 
  \item  Perhaps SN 1954J was a genuine supernova event, and the 
  bluer  SED  plus H$\alpha$ emission are associated with   
  a resulting neutron star or black hole near the G-type   
  supergiant.  If we speculate 
  that the pre-1954 state was a close binary, then the unusual 
  faintness of the SN event might be ascribed to the exploding 
  star's small size, i.e., maybe it had transferred most of its 
  mass to the G supergiant.  In some respects this is an appealing 
  scenario, but there are serious objections.  (a) The hypothetical 
  accreting neutron star of black hole should probably be an 
  X-ray source, but none has been detected \citep{Binder}.   
  (b) The supergiant should very likely have an anomalously high 
   velocity of the order of 40 km s$^{-1}$, due to its pre-event 
   orbit, but this is not evident in the spectra.  (c) This model   
   does not account for the H$\alpha$ line shape.  The asymmetry 
   of a classic Thomson-scattered profile results from expansion 
   of the gas, unlike an accretion flow toward a compact object. 
   Thomson scattering would indeed be effective in such a compact 
   H$\alpha$ source, but it would produce a different line shape.  
   (d) We don't have a model to explain why the accretion disk 
   or inflow produces a bright continuum at the blue-visual  wavelengths. 
  %%%    
  \item  Suppose that there is only one star, but \ion{Ca}{2} and other 
   absorption lines are formed in outer, cool zones of a dense wind, 
   rather than a normal photosphere.  Such an outflow would be 
   unsurprising, given the precedent of $\eta$ Car in the state 
   that was seen 60 years after its giant eruption \citep{hdk2008}.  
   Standard spectral classification criteria are not valid for 
   dense winds \citep{dh2012,kd2016}, so the star can be hotter and 
   smaller than the 5000 K supergiant deduced in Section 
   \ref{subsec:v12spectr}.  The continuum may even originate in a diffuse 
   opaque-outflow photosphere \citep{KD87,kd2016,Owocki}, rather than a 
   static stellar photosphere.   Unfortunately the dense-outflow 
   idea cannot by itself overcome the H$\alpha$ objection in point 1 
   above;  with any credible cool photosphere radius, say 
   $R > 50 \, R_\odot$, 
   one finds $\tau_\mathrm{sc} \lesssim 0.3$ in a spherical region that 
   produces the observed amount of H$\alpha$ emission.  Perhaps 
  $\tau_\mathrm{sc} \sim 1$ can be  attained with a more localized 
  emission volume -- e.g., in a bipolar flow -- but this possibility 
  is quantitatively marginal.   
  Moreover, one would expect the \ion{Ca}{2} absorption to be  
  blue-shifted relative to the H$\alpha$ peak, contrary to the data. 
  In summary, this model avoids the need for a second component 
  and it cannot be dismissed, but it appears less successful than 
  model 2 above. 
 %%%   
\end{enumerate}

None of the above scenarios is entirely satisfactory, but in our opinion 
model 2 is by far the simplest and most likely.  Except for the 
possible supernova binary model, none of them alone can explain the 
decline in apparent luminosity from $10^5 \; L_\odot$ before the eruption 
to $10^4 \; L_\odot$ afterward.  And, even in the SN model, luminosity 
indicators in the spectrum of the G-type star do not match its 
current state.  The most likely explanation for this luminosity 
discrepancy is  post-eruption dust formation and circumstellar 
extinction, {\S\S} \ref{subsec:circumst} and \ref{subsec:qmodels} below.

In the next section we examine the possibility of dust formation, and the 
limits set by the mid-infrared flux shown in the SEDs.

\subsection{Circumstellar Dust, Extinction, and Mass Loss} 
\label{subsec:circumst}  %%% ===-=== %%% 

The historical light curve  shows some evidence in support of dust 
formation.  After maximum light, the apparent brightness rapidly 
declined at blue wavelengths. By early 1957 it was fainter than the 
pre-eruption state,  and post-eruption magnitudes from 1957 to 1963 
progressively faded from 21.75 to 22.4 mag.  The few color values also 
show that it was becoming redder, consistent with increasing circumstellar 
extinction.  The faintest blue magnitude recorded in those years was 
22.4, interestingly close to what we observe today.  Of course those 
measurements referred to the sum of all four objects, but evidently one 
of them -- almost certainly star 4 -- became fainter than it had been 
a few decades earlier, and dust formation is arguably the simplest 
explanation.  Indeed it may have faded even more after 1963, followed 
by partial brightening in recent years as the dust moves outward.   
In this section  we outline the implications of dust formation in this object, 
an unsettled topic. 

Possible circumstellar dust asociated with V12 has been discussed by several
authors, without agreement.
\citet{Smith} suggested that V12 had a near-IR excess due to dust, 
based on groundbased UBVRIJHK photometry that included all 4 stars.
\citet{VanDyk05b} based their arguments for dusty ejecta on the 
H$\alpha$ line, presumably indicating a luminous hot star, together 
with star 4's location in their two-color diagram.  
They suggested a  combined  $A_{V} \approx$ 4 mag, interstellar 
plus circumstellar extinction. 
In contrast, \citet{CSK12} disputed this conclusion, arguing that the 
mid-infrared fluxes from {\it Spitzer} and obscuration could not come 
from the low luminosity and temperature that they estimate for star 4. 
The {\it Spitzer}/IRAC  and MIPS fluxes represent the integrated 
light of all four stars including the two red supergiants. Our 
improved {\it HST} photometry permits an assessment of the four objects' 
relative contributions.  

If the ejecta from V12's giant eruption did produce dust, it should 
now be at $r \gtrsim 4000$ AU with expansion speeds of 300 -- 600 
km s$^{-1}$. Assuming a luminosity of 10$^{5}$ L$_{\odot}$ for the progenitor,  
with efficiencies $Q_{\lambda} \approx$ 0.03 to 0.1 for re-radiation 
by dust, the dust temperature should be 130--180 K with peak radiation 
near $\lambda \sim $ 16--22 $\mu$m.  If the expansion was faster,  
the dust may be further from  the star, cooler and radiating at 
longer wavelengths. The {\it Spitzer}/MIPS flux in the  24$ \mu$m band 
shows a mid-infrared source, but, given the complexity of the region 
and background at such wavelengths, we consider that measurement 
to be an upper limit for V12.  With  6$\arcsec$ resolution and 
2\farcs5 per pixel,  it refers to the combined light of all four 
stars like the other infrared data. 

We can reasonably expect that hot star 3 does not make a significant 
contribution at this wavelength. The two red supergiants, however, may 
make appreciable contributions from  silicate emission around 24 $\mu$m.
They appear nearly identical in the photometry.  Using our measurement 
of the interstellar extinction, each has $M_{v} \approx -5.2$ and 
their intrinsic colors suggest an early M spectral type. The corresponding 
bolometric luminosities are $-6.4$ mag. To estimate their likely 
contribution to the 24 $\mu$m flux,  we selected four M supergiants in 
the Per OB1 association ($D \approx 2.2$ kpc) with ISO spectra and 
comparable luminosities and spectral  types.  Their average flux at 
24 $\mu$m  is 14 Jy, which scaled to the distance of NGC 2403 becomes 
$7.4 \times 10^{-6}$ Jy. Since this is far below the measured flux for 
V12, the two red supergiants therefore do not account for it. 
Although PAH emission dominates in the 8$\mu$m band, the radiation 
at 24 $\mu$m is mostly thermal emission from grains with 
temperatures $\gtrsim$ 100 K.  We therefore suggest that much of 
the mid-infrared flux in the V12 environment comes from  ejecta associated
with star 4. 

We can use the flux at 24 $\mu$m to estimate limits on the mass of the 
dusty circumstellar material and the total mass lost in the 
giant eruption.  The dust mass is given by  
 \[  
      M_{dust} = \frac{4D^{2} {\rho} {\lambda}F_{\lambda} }
                 { 3({\lambda} Q_{\lambda}/a) B_{\lambda}(T) } \; ,  
		 			     		                \] 
where D is the distance to NGC 2403, F$_{\lambda}$ is the flux, {\it a} 
is the grain radius, $\rho$ $\approx$ 3 g cm$^{-3}$ is the grain density 
for silicates, Q$_{\lambda}$ is the grain efficiency for the absorption 
and emission of radiation, and B$_{\lambda}$(T) is the Planck specific 
intensity at temperature T. For this calculation, we adopt a dust 
temperature of 150 K and $a = 0.5 \; \mu$m.    Using the formulation 
from \citet{Suh},   Q$_{\lambda} \approx 0.035$ at 21--26 $\mu$m.   
The upper limit to the  mass in dust is then 2 $\times 10^{31}$ g. 
Assuming the conventional gas to dust ratio of 100, this suggests 
that the ejected mass is on the order of 0.5 \ $M_{\odot}$.   Hence the  
average rate as V12 faded in 10 years was   
$\approx$ 0.05 $M_\odot$ y$^{-1}$. Presumably the rate during the  
event was much higher, and perhaps unsteady.  If some of the  
dust is at large optical depths, i.e., shielded from most of 
the star's light, then it may be cool and thus excluded from  
the above estimate. There also may have been additional mass loss during
V12's erratic short term  outbursts prior to maximum. 

If circumstellar dust absorbs most of the light from star 4, then 
the 24 $\mu$m data may constrain the total luminosity.  The reasoning, 
however, is far from robust.  If some of the 24 $\mu$m flux comes 
from unrelated dust in a larger region, then this method overestimates 
the amount of light from star 4;  or, on the other hand, if some of 
the light escapes through gaps in the dust shell, then we get an   
underestimate.  Moreover, we know very little about the grain  
properties, and $\eta$ Car's unusual grains serve as a cautionary  
example regarding eruptions \citep{KD97}.  Nevertheless, we adopt  
simplified assumptions here to estimate the dust luminosity. 
If the grain radiative efficiency has wavelength dependence 
$Q_\mathrm{abs} \propto {\lambda}^{-2}$ in the mid-infrared, and 
the 24 $\mu$m flux approximates the peak of the $\lambda f_{\lambda}$ 
curve, then the grain temperature is about 100 K.  In that case  
the observed $\lambda F_\lambda \approx 2.6 \times 10^{-16}$ W m$^{-2}$ 
implies a total flux $F_\mathrm{IR} \approx 4 \times 10^{-16}$ W m$^{-2}$ 
and $L_\mathrm{IR} \approx  10^{5} \; L_\odot$ at NGC 2403.\footnote   
   %%%% FOOTNOTE
      {Here we include a 30\% allowance for the fact that the dust SED 
         is broader than the Planck function, since it includes a range  
	    of temperatures.  }  
	       %%%%%
If instead the dust temperature is 150 K, then the peak $L_\mathrm{IR}$ 
is 50\% larger.   If $Q_\mathrm{abs} \propto {\lambda}^{-1}$,   
then the optimum grain temperature is about 120 K instead of 100 K 
and $L_\mathrm{IR}$ is about 10\% above the other estimates.  
In addition, the 8 $\mu$m  upper limit implies that $L_\mathrm{IR}$ cannot 
be larger than about twice the quoted limit.  Allowing for some of 
the uncertainties, $L \sim 2 \times 10^{5} \; L_\odot$ appears 
to be a credible upper limit -- except that the reasoning becomes invalid 
if a substantial fraction of the light escapes through gaps in the 
dust shell. 

Circumstellar dust can evidently account for most of  the apparent 
post-eruption decrease in the brightness noted above.   Star 4's 
current luminosity corrected for interstellar (not circumstellar) 
extinction is roughly $10^4 \; L_\odot$.  The various quantities 
appear to permit a circumstellar extinction factor of order 10, 
thereby allowing a corrected $L \sim 10^5 \; L_\odot$ which one 
informally expects for a giant eruption survivor.  This question 
is discussed further in {\S} \ref{subsec:qmodels} below.  

Extinction of the order of two magnitudes normally implies conspicuous 
reddening,  but this depends on grain properties which can be peculiar 
in a giant eruption or massive stellar wind.  Anomalous 
extinction is well documented for, e.g., stars in the Carina Nebula 
\citep{The}. The extinction/reddening 
ratio $A_V/E_{B-V}$ ranges from a normal value near 3 to a high of 6,  
and probably higher for $\eta$ Car.  In other words, strong extinction 
does not always imply strong reddening.  Anomalous extinction curves 
of this type are attributed to large grain sizes, and play a role 
in the next subsection.  

\subsection{A Quantitative Model with Circumstellar Reddening} 
\label{subsec:qmodels} %%%%%%%

As noted above, two strong clues imply the presence of circumstellar 
extinction for star 4 (which is  two objects). 
First, if we apply 
only the interstellar correction used in Figs.\  \ref{fig:sed4} and 
\ref{fig:2comp}, the 5000 K supergiant at 0.8 $\mu$m  appears to have 
$L \sim 4000$ to $8000 \; L_\odot$ or $M_\mathrm{bol} \approx$ -4.6 to -5.
But the luminosity indicators in its spectrum imply a much higher 
luminosity ({\S} \ref{subsec:v12spectr}), thus 
requiring an additional extinction correction of roughly 2 to 3 
magnitudes at $\mu \approx 0.8 \ {\mu}$m.      
Secondly, note the pre-eruption brightness indicated in the upper-left 
part of Fig.\ \ref{fig:sed4}.  As  explained in 
{\S} \ref{subsec:v12extinct}, it represents only  star 4, the others 
being subtracted out.  The present-day star 4 appears fainter by 
roughly 2.8 magnitudes in the `B' band.  Thus, if it is in 
the same state as in 1910--1938, we need that amount of additional 
extinction near $\lambda \approx 0.45 \ \mu$m.  
(Later we mention physical reasons why the eruption may have caused 
a decrease in brightness at blue wavelengths, but they cannot 
explain a factor larger than about 0.8 magnitude.)  We assume 
that the circumstellar extinction equally affects the hot component, 
and the G supergiant, and the H$\alpha$ emission;  otherwise a 
self-consistent model becomes more difficult.  

\begin{figure}
\figurenum{8}
\epsscale{1.2}
\plotone{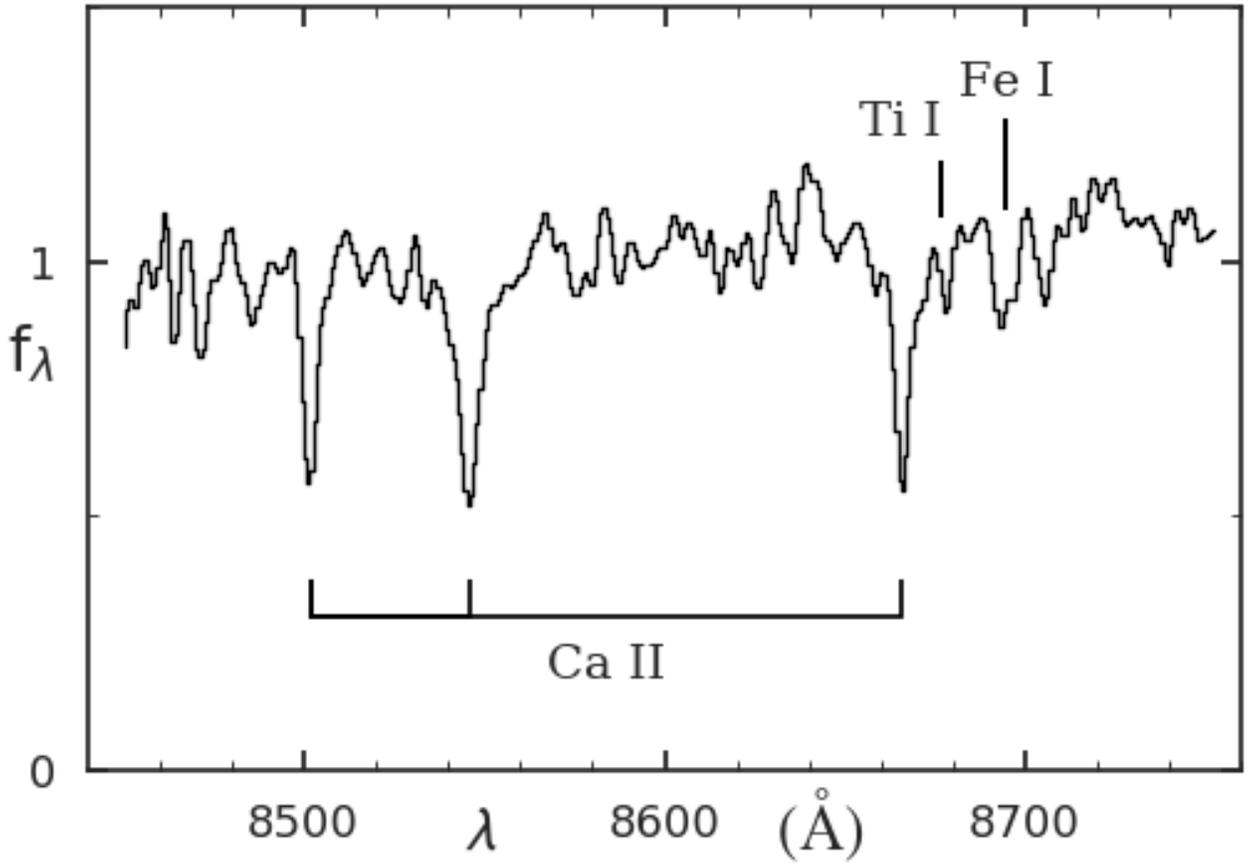}    
\caption{The Ca II triplet in V12. The absorption lines of Ti I($\lambda$ 
8675{\AA}) and Fe I($\lambda$ 8688{\AA}) are also identified.  
The units are  10$^{-17}$ ergs cm$^{-2}$ s$^{-1}$ A$^{-1}$.}
\label{fig:caii}  	
\end{figure}

\begin{figure}
\figurenum{9}
\epsscale{1.0}
\plotone{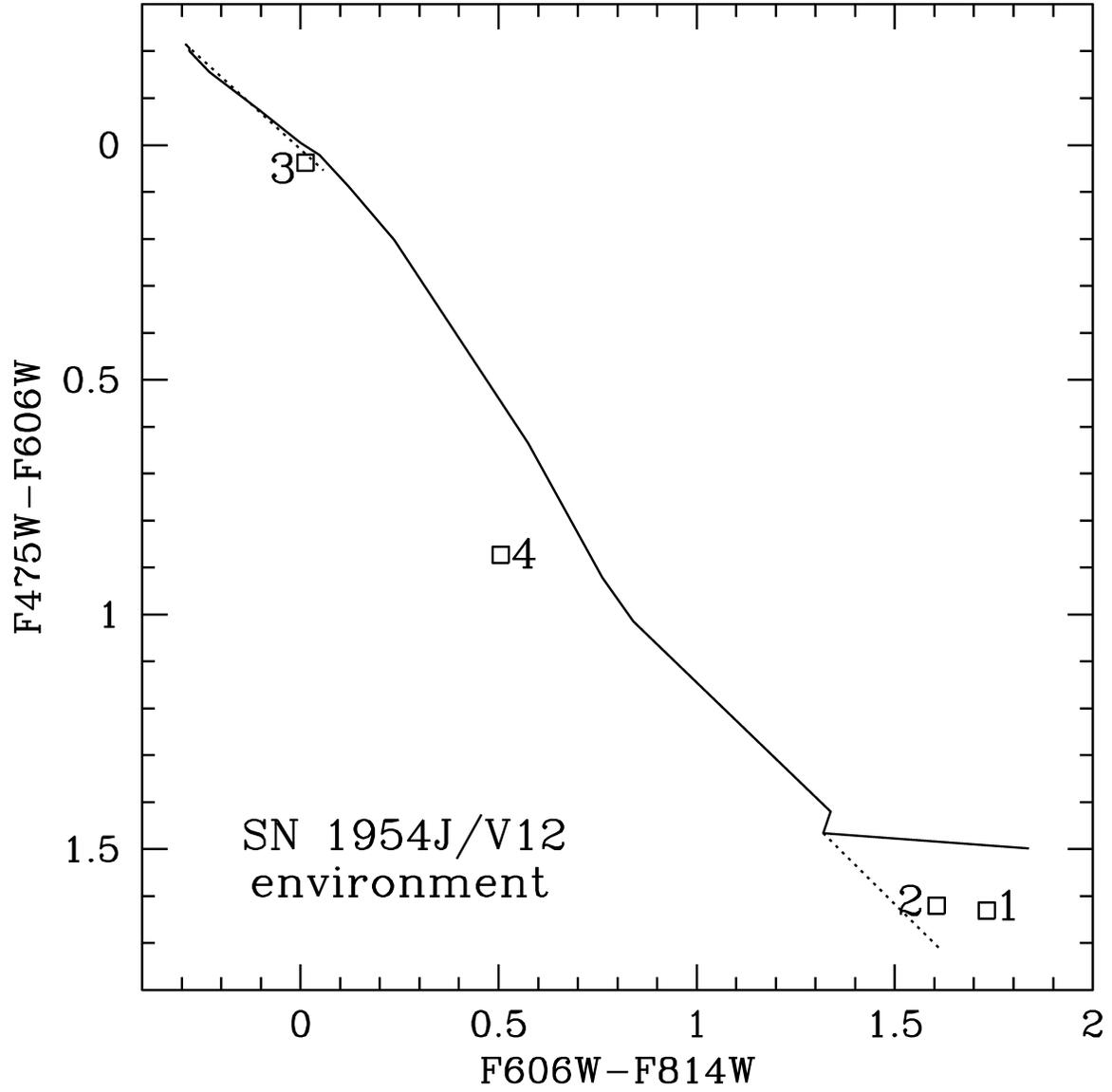}
\caption{The two-color diagram for the four stars in the V12 environment 
    using the VEGAMAGS from 2004.}
\label{fig:2color}	
\end{figure}

\begin{figure}
\figurenum{10a}
\epsscale{1.0}
\plotone{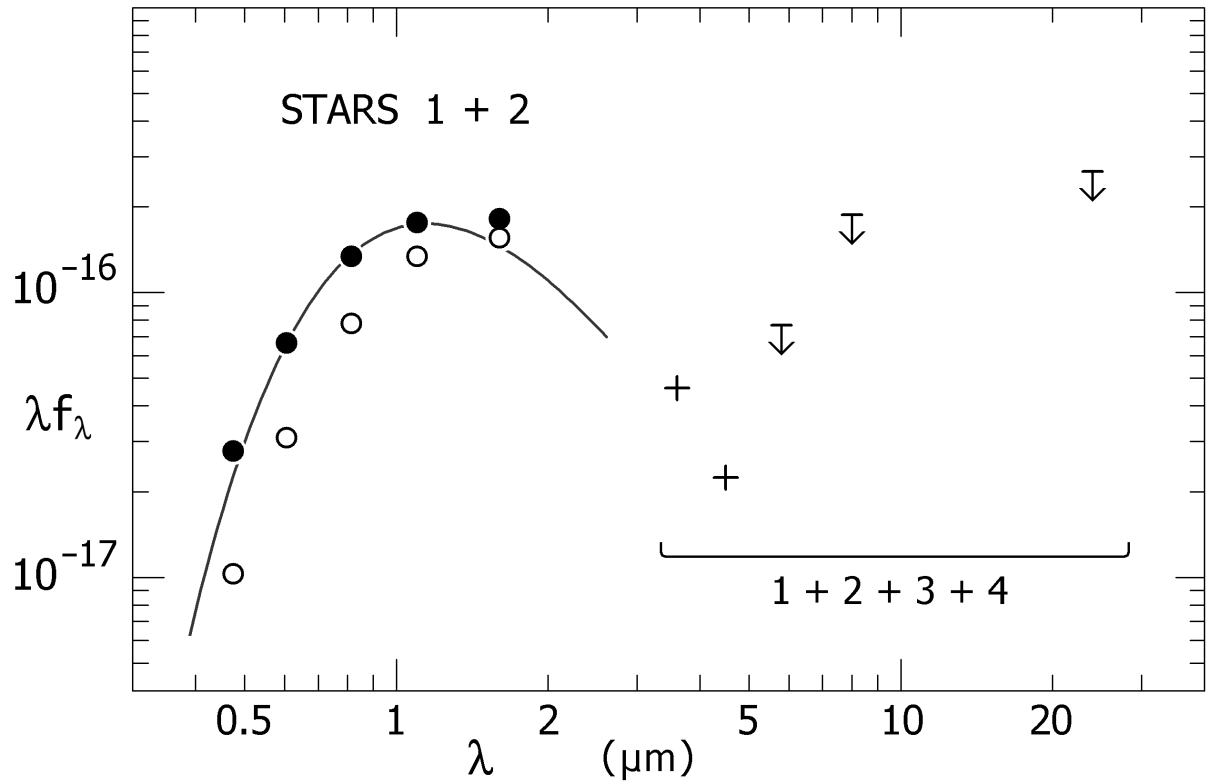}
\caption{The combined SED for the two red stars.  A 3200 K Planck curve is 
   shown for their extinction-corrected combined fluxes. Open circles are 
   the observed fluxes based on VEGAMAGs and the filled circles are corrected 
   for interstellar extinction, A$_{V}$ = 0.8 mag.  These two red stars are 
   the probable source of near-IR flux at the position of V12.  Units  
   for $\lambda f_\lambda$ are W m$^{-2}$. }
\label{fig:sed1and2} 	
\end{figure}

\begin{figure}
\figurenum{10b}
\epsscale{1.0}
\plotone{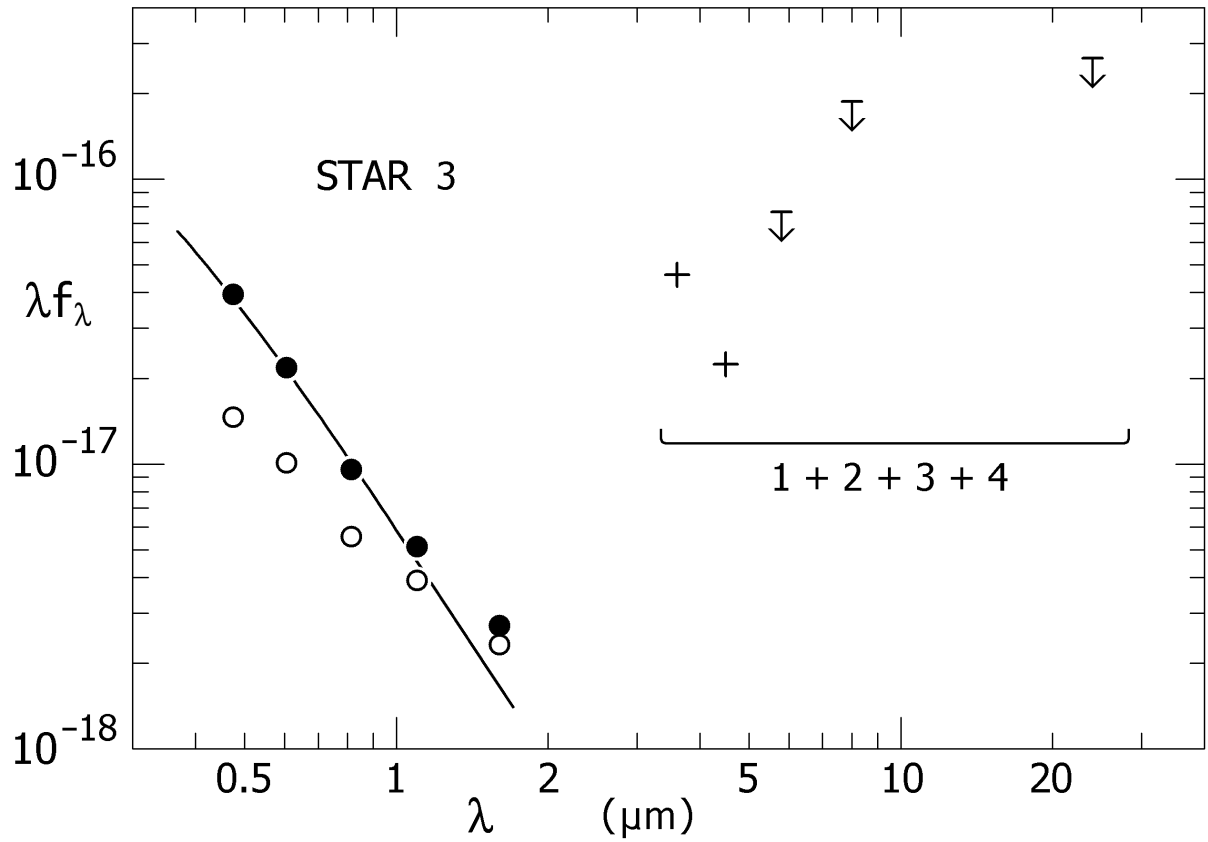}
\caption{The SED for star 3. Open circles are observed fluxes based on 
   VEGAMAGs, and filled circles are corrrected for interstellar extinction, 
   A$_{V}$ = 0.9 mag.  A 25,000 K Planck curve is shown.  
	 $\lambda f_\lambda$ is expressed in W m$^{-2}$. }
\label{fig:sed3}
\end{figure}

\begin{figure}
\figurenum{10c}
\epsscale{1.0}
\plotone{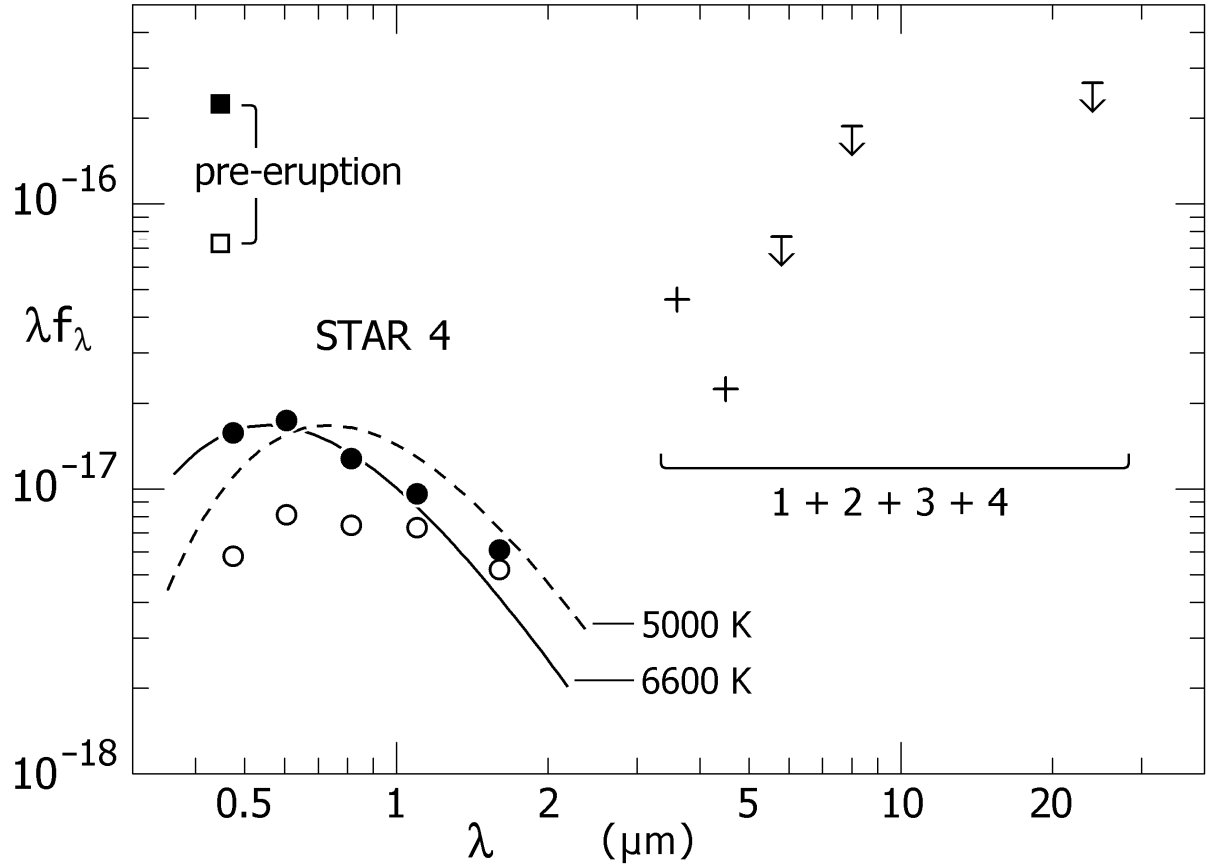}
\caption{The SED for star 4.  Observed fluxes from VEGAMAGs are  shown 
   as open circles, and filled circles are corrected for interstellar 
   extinction A$_{V}$ = 0.9 mag.   The best-fit Planck curve has  
   $T = 6600$ K, but the absorption line spectrum indicates 5000 K. 
   The pre-eruption magnitude in 1910--1938 is also shown.  
   Units for  $\lambda f_\lambda$ are  W m$^{-2}$. }
\label{fig:sed4}  
\end{figure}

\begin{figure}
\figurenum{11}
\epsscale{1.0}
\plotone{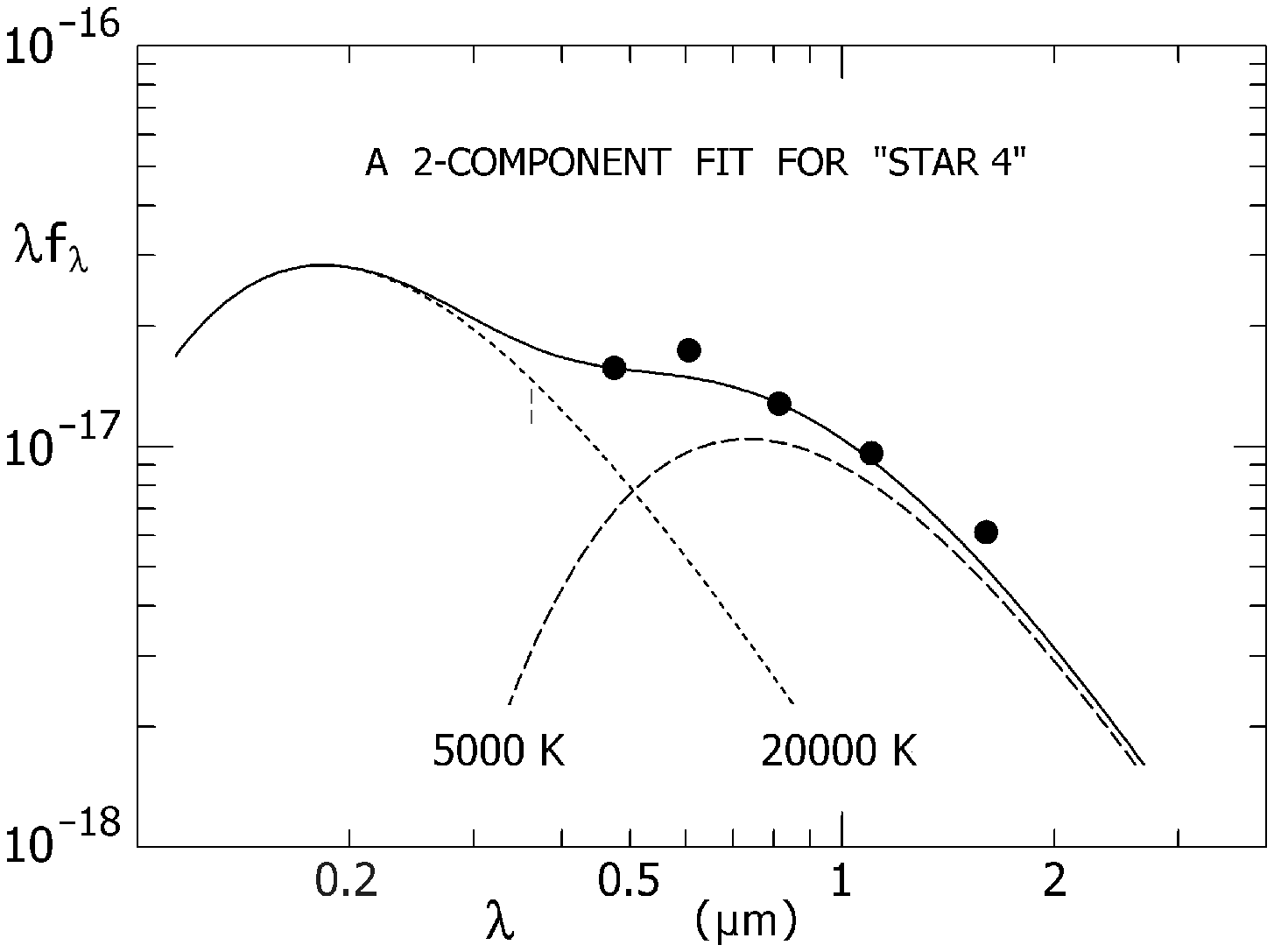}
\caption{A typical two-component fit to the SED for star 4 with a 5000 K Planck 
curve inferred from the G-type absorption lines and a hotter 20,000 K star as 
discussed in Sections 4.4 and 4.6. A small vertical mark indicates the Balmer 
discontinuity wavelength. Possible circumstellar extinction is not included.} 
\label{fig:2comp}
\end{figure}

The above clues favor circumstellar extinction with $A_V \sim 2.5$ 
and only  weak reddening, $A_B - A_V < 0.5$.   
Moreover, stronger reddening would almost preclude a consistent  
two-star model.  In such a case the hotter star is found   
to dominate at all wavelengths, making it impossible for the  
\ion{Ca}{2} triplet  and the other neutral lines to be strong as 
they are in the combined spectrum (Fig.\ \ref{fig:caii}).  
Hence there are two independent 
reasons to suppose that the reddening is weak.  This is quite 
reasonable, since large extinction/reddening ratios 
are  often associated with very massive stars as mentioned in 
{\S} \ref{subsec:circumst}  An extreme example is $\eta$ Car, with 
high circumstellar extinction but only imperceptible reddening 
\citep{KD97}.  Apparently the outflows from some massive stars 
produce unusually large grains, and, given SN 1954J's eruption, 
we should not be surprised if it belongs in that category.

The data are not adequate for a unique model, but here is a   
nearly satisfactory example.  Suppose that the two stars have 
color temperatures $T_1 = 5000$ K and $T_2 = 20000$ K;  the former 
is indicated by the absorption lines while the latter is more 
arbitrary, but a much hotter member 
is not supported by the spectrum, \S {4.4}\footnote{Note that these are not 
effective temperatures. If the hot star has a Balmer jump, then its "effective" temperature may be somewhat less. However we do not have sufficient data to quantify this detail.}. 
 Further assume that the circumstellar 
extinction is 2.5 and 2.8 magnitudes at $\lambda = 0.8 \ \mu$m      
and $0.45 \ \mu$m respectively, and this applies equally to 
both stars and to the H$\alpha$ emission.  Then, by matching   
the corrected fluxes at 0.45 $\mu$m and 0.8 $\mu$m, 
we find the following results.  The cool star has luminosity  
$L_1 \approx 3.6 \times 10^4 \ L_\odot$, 
$M_\mathrm{bol} \approx -6.7$, and radius 
$R_1 \approx 250 \; R_\odot$,  while its hot   
companion has $L_2 \approx 2 \times 10^5 \ L_\odot$,  
$M_\mathrm{bol} \approx -8.5$, and $R_2 \approx 37 \; R_\odot$.  
The hot star contributes only about 30\% of the continuum 
at $\lambda \approx 0.85 \; \mu$m near the \ion{Ca}{2} triplet 
(Fig.\ \ref{fig:caii}).  Its parameters appear   
satisfactory for an evolved massive star with mass loss and a continuing 
dense wind.   Based on stellar models \citep{Ekstrom} they correspond to initial masses for evolved stars of
$\sim 20 M_{\odot}$ and $\sim 15 M_{\odot}$ for the hot and cool star, respectively.

We do not know if this is a physical pair, but the G supergiant has been
affected by some additional extinction, and therefore must be within the
dust cloud, 4000 - 8000 AU from the hot star. Based on stellar models with rotation \citep{Ekstrom}, a post-red supergiant of 20 $M_\odot$ near the end of its track, will be 10 million  years old while the somewhat lower mass G supergiant,
$\sim$ 15 $M_\odot$, will be $\sim$ 11-13  million years old, without or with 
rotation, respectively, in the stellar models. 
Given the uncertainties in the above 
model and the derived parameters, this is reasonable agreement for a possible pair.

The 20000 K star is also suitable for producing the H$\alpha$ 
feature.  Given the hot star's radius, the observed 
$L(H\alpha \approx 1.4 \times 10^{37}$ ergs s$^{-1}$) can be 
emitted by a dense wind with density 
$n_e \sim 4 \times 10^{11}$ cm$^{-3}$ just outside the photosphere. 
The corresponding Thomson-scattering depth is 
$\tau_\mathrm{sc} \approx 0.7$, which is nearly adequate for 
explaining the observed profile, given the various uncertainties.     
(Note that the H$\alpha$ line played no role in calculating the 
star's radius.)    
Admittedly we have not corrected for inhomogeneous ``clumping'' 
in the wind, but on the other hand it may be bipolar or equatorial, 
which can increase the relevant size of $\tau_\mathrm{sc}$.   If the 
outflow velocity is  300 km s$^{-1}$, then the implied mass loss 
rate for the stellar wind is of the order of 
$3 \times 10^{-5} \ M_\odot$ y$^{-1}$.   

Strictly speaking, the above model slightly violates an 
upper luminosity limit described in {\S} \ref{subsec:circumst}, 
based on the 24 $\mu$m flux from dust.  But this is not a very 
serious objection, since many of the assumptions can easily 
be readjusted.  For instance, the 24 $\mu$m argument fails if 
some light   escapes through gaps in the dust.  And the 
circumstellar extinction may be somewhat smaller than we have 
assumed, because the present-day hot star may be intrinsically 
somewhat fainter than it was before the eruption.  (Having lost 
some mass and also some energy, it can now be slightly less 
luminous and appreciably hotter than it was then.)  
Given these uncertainties, the model described above confirms 
that a two-component interpretation of star 4 is feasible.  
Moreover, it is less arbitrary that the other concepts   
outlined in {\S} \ref{subsec:models} above. 
Pending new evidence, we conclude that {\it the survivor 
of SN 1954J is most likely an evolved  blue supergiant, with a companion 
G-type supergiant.\/}  The blue supergiant most likely had     
an initial mass around 20 $M_\odot$. We suspect that it is a post-red
supergiant, on a blue loop, and 
has lost about half of its initial mass  so that its proximity to 
the Eddington Limit will reduce its stability \citep{HD94,kd2016}.   

This result is practically what one would expect for a giant eruption 
\citep{HD94}. It qualitatively resembles Van Dyk's (Van Dyk et al.(2005b) 
suggestion based on the same expectations from circumstellar dust. 
Kochanek et al. (2012) mentioned similar parameters for the likely 
progenitor but rejected it based on the dust properties. 

{\it A good signal to noise blue spectrum  of this faint star is needed 
to further refine the 
physical characteristics of the likely hot star progenitor and to better understand 
its  evolutionary state.}

\section{Summary and Discussion}
%%\subsection{sec:summary}   %%% ===-=== %%%

Our two supernova impostors, SN 2002kg and SN1954J, are survivors. Each was the eruption or high mass loss event of an evolved massive star, but they have  very different stories to tell.  

SN2002kg or V37 is an  LBV/S Dor-type variable. In 2002 it had a high mass 
loss event and was initially classified as a Type IIn supernova with narrow hydrogen 
lines in emission.  However, it was sub-luminous and was identified with V37 in NGC 2403 \citep{Weis05,VanDyk05a}, a known ``irregular blue variable''\citep{ST1968}. 
Its pre-eruption  
colors show that V37  was a hot O9-B0 star. During its eruption,  V37 formed a cool, dense wind with absorption lines typical of  an early A-type supergiant in its spectra near maximum light, in addition to the strong, narrow hydrogen emission.  
Its light curve shows that its high mass loss stage lasted 
about 13 years and it 
has nearly returned to its pre-eruption or quiescent magnitude. A current spectrum now shows a much hotter star, as expected, with He I emission and P Cygni profiles 
due to its stellar wind. V37  also has a circumstellar nebula which very likely 
formed during the recent eruption, although it  did not form dust.
We show that V37's pre-eruption luminosity and its luminosity near maximum  was 
approximately M$_{Bol}$ -10.4 mag or 10$^{6}$ L$_{\odot}$ and probable initial mass of 60 - 80 M$_{\odot}$.  
Like other LBVs/S Dor variables, V37s outburst was an apparent transit on the 
HR Diagram, at nearly constant luminosity due to increased mass loss and the 
formation of a cooler, dense wind. With its luminosity and apparent temperature
of $\approx$ 30,000 K from its intrinsic colors, V37 will lie on the 
LBV/S Dor instability strip \citep{Wolf,HD94} near classic LBVs such as R127 in the LMC.

 Recently \citet{Smith15}  suggested that LBVs are ``kicked mass gainers,'' 
i.e.,  former binary components that gained mass from   
pre-supernova companions and are now runaways.  Their arguments 
were based on statistical isolation of LBVs from other massive stars, specifically O-type stars.  \citet{RMH2016} found, however, that genuine luminous 
``classical LBVs'' 
are {\it not\/} isolated, while the less luminous LBVs differ in 
this respect because they are substantially older.  Altogether 
the spatial distributions agree with standard expectations. 
(\citet{Smith2016} disputed these findings, but \citet{DH2016} listed 
specific invalid assumptions  in each of his objections.)  The most important 
outcome was strong arguments favoring the standard view of LBVs as evolved massive stars. 
How do V12 and  V37 fit into this context?  
First, V37 is not isolated.  It is in a spiral arm  and associated with
other reddened hot stars \citep{VanDyk06,Maund06}, similar to the classical LBVs in our Local Group galaxies \citep{RMH2016}.
 V12 is not an LBV/S Dor variable. It was a giant eruption.  Although it is a probable binary, there is no reason to require mass exchange as in the \citet{Smith15} model for LBVs. Furthermore, V12 is not isolated. It is in a spiral arm and its immediate
 environment (Figure 1) includes a hot supergiant, two red supergiants, and a G-type supergiant companion.

In its  giant eruption, SN1954J/V12 increased by five magnitudes in 
apparent brightness and its absolute magnitude at maximum was at least
$-$ 13 or $\approx$ 10$^{7}$ L$_{\odot}$. Its maximum luminosity and  its 
pre-maximum outbursts are similar to other SN impostors or giant eruptions. 

Based on V12's  pre-eruption light curve we estimate an M$_{B}$ of -7.2 for 
V12's progenitor, corrected for the interstellar extinction in its environment.  With no color information at that time we cannot directly estimate its 
temperature and luminosity. Assuming the approximate $B-V$ color of 1.0 
observed during the rise to maximum, its luminosity was $\approx$ 10$^{5}$ L$_{\odot}$  with an initial mass of 20 M$_{\odot}$. The likely survivor, 
the H$\alpha$ bright star 4 in the 
{\it HST} images, however is now  2.5 to 3  magnitudes fainter than 
the pre-eruption progenitor with a luminosity of only $\approx$  10$^{4}
L_{\odot}$. 

V12/star 4's red spectrum  shows prominent H$\alpha$ emission with an electron scattering stellar wind profile plus the Ca II triplet in absorption and 
neutral metallic lines of Fe I and Ti I.  The absorption lines are consistent 
with a late G-type supergiant with a temperature of about 5000 K. V12's 
colors, however are anomalous; too blue for a G-type star. Its SED is also peculiar and cannot be fit by a single 5000 K star. We therefore suggest that 
star 4 is actually 
two stars; a hotter star responsible for the H$\alpha$ emission plus the G-type supergiant. Star 4's SED can be adequately fit by a two component model
with a  
hotter star in the range temperature 15000 to 25000 K plus the 5000 K cooler star. 

We also demonstrate that if the 24$\mu$m flux in the V12 environment is the 
remnant radiation from V12's ejecta,  the total mass shed in the eruption was 
$\sim$ 0.5 M$_{\odot}$. The 24$\mu$m flux also sets constraints on the 
luminosity radiated by the dust, $L_\mathrm{IR} \sim  10^{5} \; L_\odot$ comparable to most of the flux estimated for the progenitor star. Re-radiation by dust can thus account for most of  
the apparent post-eruption decrease in the brightness  of V12.  

With additional circumstellar extinction, the hotter component, assuming 
a 20,000 K star, would have a luminosity of 2 $\times$ 10$^{5}$ L$_{\odot}$ comparable to that originally estimated for the progenitor star. If the G supergiant also suffers circmstellar extinction its corresponding luminosity would be $\approx$
3.6  $\times$ 10$^{4}$ L$_{\odot}$.

Our analysis of these two impostors has revealed 
two very different stars. V37/SN2002kg is a classical LBV, an evolved massive 
star, 60 - 80 M$_{\odot}$, experiencing enhanced mass loss episodes due to its 
proximity to its Eddington limit \citep{RMH2016}. Curiously, the much lower mass V12/SN1954J had the giant eruption, but its high mass loss event was also 
fueled by its proximity to its Eddington limit. Based on  our estimates for the temprature and luminosity of the hot star, V12 would lie 
just below the LBV instability strip on the HR Diagram. As has been 
suggested for the ``less luminous'' LBVs \citep{HD94,Vink2012}, V12 was very 
likely  a post-red supergiant that had evolved back to warmer temperatures. 
As a red supergiant, it would have lost a lot of mass bringing it closer to 
its Eddington limit and more subject to interior as well as surface 
instabilities. 
Stellar models \citep{Ekstrom} with mass loss and rotation for  
20 M$_{\odot}$ and 25 M$_{\odot}$ show that these stars, on a blue-loop,  
would have shed 
half their initial masses when they reach the blue side of the HR Diagram. 
Interestingly, near the end of the model, the 
20 M$_{\odot}$  track shows a short transit to the red, an ideal 
state for a star to experience a high mass loss event as it tries to evolve
 to cooler temperatures.  
Finally, if the interpretation of the plateau on its declining light curve as a collision with  previous
ejecta is correct,  then V12/SN1954J had experienced high mass events, even giant eruptions, in the past.

Research by R. Humphreys and K. Davidson on massive stars is supported by  
the National Science Foundation AST-1109394.
This paper uses data from the MODS1 and MODS2 spectrographs built with funding from NSF grant 
AST-9987045 and the NSF Telescope System Instrumentation Program (TSIP), with
 additional funds from the Ohio Board of Regents and the Ohio State University Office 
 of Research.

{\it Facilities:} \facility{MMT/Hectospec,LBT/MODS1 + MODS2, HST/ACS, HST/WFC3}

%%%%%%%%Tables

\begin{deluxetable}{llllllllllllll}
\rotate
\tablewidth{0 pt}
\tabletypesize{\footnotesize}
\tablenum{1} 
\tablecaption{Observed Colors and Luminosities for V37} 
\tablehead{
\colhead{Date} & \colhead{V} & \colhead{B-V} & \colhead{U-B} & \colhead{V-R} & \colhead{V-I} &
\colhead{E(B-V)} & \colhead{V$_{0}$} & \colhead{B-V$_{0}$} & \colhead{U-B$_{0}$} & \colhead{V-R$_{0}$} & \colhead{V-I$_{0}$} & \colhead{M$_{v}$} & \colhead{Notes} 
}
\startdata
13 October 1997 & 20.63 & $-$0.08  & $-$0.96 & 0.09 & $-$0.06 & 0.19 & 20.09 & $-$0.25 & $-$1.09 & $-$0.05 & $-$0.34 & -7.4 & pre-max. \\
20 February 2001 & 20.65 & 0.27 & \nodata & \nodata & $-$0.29 & 0.17 & 20.11  & \nodata & \nodata & \nodata & \nodata & -7.4 & pre-max \\
26 March 2003 & 18.32 &  0.13  & \nodata & 0.28  &  0.35 &  \nodata & 17.78 & -0.04 & \nodata & 0.14 & 0.07 & -9.8 & near max.\\
17 August 2004\tablenotemark{a} & 19.43 & 0.21 & \nodata & \nodata & 0.21 &  \nodata & 18.89 & 0.04 & \nodata & \nodata & -0.07 & -8.6 & post-max.\\ 
\enddata
\tablenotetext{a}{Revised ACS/WFC  photometry using DOLPHOT. Additional B - and R-band magnitudes were measured on 21 September 2004 with the ACS/HRC; B = 19.54, R = 19.24.} 
\end{deluxetable}

\begin{deluxetable}{lllllllll}
\rotate
\tablewidth{0 pt}
\tabletypesize{\footnotesize}
\tablenum{2}
\tablecaption{Multicolor Photometry for the Four Stars in the V12 Environment}
\tablehead{
\colhead{Star}  &  \colhead{F475W} & \colhead{F606W} & \colhead{F814W} & 
\colhead{F110W} &  \colhead{F160W}  &   \colhead{B}  &  \colhead{V}  &  
\colhead{I} 
  }

\startdata
August 17, 2004(ACS/WFC) &    &   &    &    &    &   &   &  \\
 1       & 24.22 &  22.59 &  20.86  & \nodata & \nodata & 24.96 & 23.10 & 20.87\\
 2       & 24.22 &  22.60 &  20.99  & \nodata & \nodata & 24.96 & 23.10 & 21.00 \\
 3       & 23.09 &  23.06 &  23.04  & \nodata & \nodata & 23.08 & 23.06 & 23.04\\
 4       & 24.11 &  23.23(23.32)\tablenotemark{a} &  22.73  & \nodata & \nodata & 24.50 & 23.44(23.52)\tablenotemark{a} & 22.72\\
         &       &        &         &         &         &       &       &      \\
October 15, 2013(WFC3)	 &    &    &  &    &       &       &      \\
 1       & 24.24 &  \nodata & 20.83 &  19.55  &  18.58  & \nodata & \nodata & \nodata\\
 2       & 24.15 &  \nodata & 20.90 &  19.68  &  18.71  & \nodata & \nodata & \nodata\\
 3       & 23.11 & \nodata &  23.03 &  22.70  &  22.46  & \nodata & \nodata & \nodata\\
 4       & 24.11 & \nodata &  22.70 &  22.02  &  21.58  & \nodata & \nodata & \nodata\\
\enddata
\tablenotetext{a}{Corrected for H$\alpha$.}
\end{deluxetable} 

\begin{deluxetable}{llllll}
%\rotate
\tablewidth{0 pt}
\tabletypesize{\footnotesize}
\tablenum{3}
\tablecaption{{\it Spitzer}/IRAC and MIPS Fluxes($\mu$Jy)} 
\tablehead{
\colhead{Star}  &  \colhead{3.6$\mu$m} & \colhead{4.5$\mu$m} & \colhead{5.8$\mu$m} &
\colhead{8.0$\mu$m} & \colhead{24$\mu$m}
}
\startdata
SN2002kg/V37 & $<$ 17  &  $<$ 14  &  $<$ 10 & $< $30 & $<$750\\
SN1954J/V12  & 54.6 $\pm$ 0.4 & 33.2 $\pm$ 0.5 & $<$ 147 & $<$ 493 & $<$ 2100$\pm$ 15 \\
\enddata
\end{deluxetable}

%%%%%%%%%%%%%%%%%Figures

\begin{thebibliography}{}
%% 
\bibitem[Auer \& Van Blerkom(1972)]{auervanb}Auer, L.H., \& Van Blerkom, 
 \apj, 178, 175
%%  
\bibitem[Binder etal.(2015)]{Binder}Binder, B., et al. 2015, \aj, 150, 94
%%
%%\bibitem[Bond(2011)]{Bond}Bond, H. E. 2011, \apj, 737, 17 
%% 
\bibitem[Cardelli et al.(1989)]{Cardelli}Cardelli, J. A., Clayton, G. C., \& 
  Mathis, J. S. 1989, \apj, 345, 245 
%%  
\bibitem[Castelli \& Kurucz(2004)]{Castelli}Castelli, F.\&  Kurucz, R. L. 2004,
  arXiv:astro-ph/0405087 
%%
\bibitem[Cenarro et al.(2001)]{Cenarro}Cenarro, A. J., Cardiel, N., Gorgas, J., Peletier
 , R. F., Vazdekis, A., \& Prada, F. 2001, \mnras, 326, 959
%%
\bibitem[Davidson(1987)]{KD87}Davidson, K. 1987, \apj, 317, 760 
 %%  
\bibitem[Davidson et al(1995)]{kd95}Davidson, K., et al. 1995, \aj, 109, 1784 
 %%  
\bibitem[Davidson \& Humphreys(1997)]{KD97}Davidson, K. \& Humphreys, R. M. 1997,
    Ann. Rev. Astronomy \& Astrophysics, 35, 1 
%% 
\bibitem[Davidson \& Humphreys(2012)]{dh2012}Davidson, K., \& Humphreys, R. M. 
    2012, Nature, 486, E1 
%% 
\bibitem[Davidson(2016)]{kd2016}Davidson, K. 2016, JphCS, 728, 2, 022008  
%%
\bibitem[Davidson et al.(2016)]{DH2016}Davidson, K., Humphreys, R. M., \& Weis, K. 2016, arXiv:1608.02007 
%%  
\bibitem[Dessart et al.(2009)]{dessart}Dessart, L, et al. 2009, \mnras, 394, 21
%%  
\bibitem[Dolphin(2000)]{Dolphin}Dolphin, A. E. 2000, \pasp, 112, 1383
%%  
\bibitem[Eck et al.(2002)]{Eck}Eck, C. R., Cowan, J. J., \& Branch, D. 2002, \apj, 57 3, 306 
%%
\bibitem[Ekstr\"{o}m  et al.(2012)]{Ekstrom}Ekstr\"{o}m, S., Georgy, C., 
Eggenberger, P., et al.  2012, \aap, 537, 146 
%%
\bibitem[Fabricant et al(1998)]{Fab98}Fabricant, D. G., Hertz, E. N., 
  \& Szentgyorgyi, A.  H., et al. 1998, Proc. SPIE, 3355, 285
%%  
\bibitem[Fabricant et al(2005)]{Fab05}Fabricant, D. et al. 2005, \pasp, 117, 1411
%%  
%\bibitem[Flower(1977)]{Flower75}Flower, P.J. 1977, \aap, 54, 31 
%% 
\bibitem[Filippenko \& Chornock(2003)]{Filip03}Filippenko, A. V. \& Chornock, R. 
 2003, IAU Circ. 8051
%%  
\bibitem[Flower(1996)]{Flower96}Flower, P.J. 1996, \apj, 469, 355   
%%  
%%\bibitem[Foley et al(2011)]{Foley}Foley, R. et al. 2011, \apj, 732, 32 
%%  
%%\bibitem[Fraternali et al.(2002)]{Fraternali}Fraternali, F, van Moorsel, G., Sancis
%   i, R. \& Oosterloo, T. 2002, \aj, 123, 3124 
 %%
\bibitem[Freedman \& Madore(1988)]{Freedman}Freedman, W. L. \& Madore, B. F. 
   1988, \apj, 332, 63 
%%  
\bibitem[Gerke(2013)]{Gerke}Gerke, J. 2013, private communication 
%%
\bibitem[Gordon et al.(2016)]{Gordon}Gordon, M. S., Humphreys, R. M., \& Jones, T. J. 2016, \apj, 826, 50  
\bibitem[Humphreys \& Davidson(1994)]{HD94}Humphreys. R. M. \&  Davidson, K. 1994, 
 \pasp, 106, 1025 
%%  
\bibitem[Humphreys et al.(1999)]{RMH99}Humphreys. R. M., Davidson, K, \& Smith, N.
  1999, \pasp, 111, 1124 
%%  
\bibitem[Humphreys et al.(2008)]{hdk2008} Humphreys, R. M., Davidson, K., \& 
  Koppelman, M. 2008, \aj, 135, 1249   
%%  
\bibitem[Humphreys et al.(2012)]{HD2012}Humphreys. R. M., Davidson, K, Jones, T. J., 
 Pogge, R. W., Grammer, S. H., Prieto, J. L.. Pritchard, T. A. 2012, \apj, 760, 93 
%%
\bibitem[Humphreys et al.(2013)]{RMH13}Humphreys, R. M., Davidson, K., Grammer, S. et al. 2013, \apj, 773, 46 
%%
\bibitem[Humphreys et al.(2014a)]{RMH14a}Humphreys, R. M. Davidson, K, Gordon, M. S.,
   Weis, K, Burggraf, B., Bomans, D.~J., \& Martin, J. C. 2014a, \apjl, 782, L21 
%%
\bibitem[Humphreys et al.(2014b)]{RMH14b}Humphreys. R. M., Weis, K., Davidson, K. 
  \& Bomans, D. 2014b, \apj, 790, 48
 %%
\bibitem[Humphreys et al.(2016)]{RMH2016}Humphreys, R. M., Weis, K., Davidson, 
K., \& Gordon, M. S. 2016, \apj, 825, 64 
%%
\bibitem[Keenan \& Hynek(1945)]{Keenan}Keenan, P. C. \& Hyneks, J. A. 1945, \apj, 101
  , 265
%%  
\bibitem[Kochanek et al.(2011)]{CSK11}Kochanek, C. S., Szczygiel, D. M. \& 
   Stanek, K. Z. 2011, \apj, 737, 76 
%%  
\bibitem[Kochanek et al.(2012)]{CSK12}Kochanek, C. S., Szczygiel, D. M. \& 
   Stanek, K. Z. 2012, \apj, 758, 142 
%%  
\bibitem[Kowal et al.(1972)]{Kowal}Kowal, C. T. et al. 1972, \pasp, 84, 844 
%%  
\bibitem[Kraus et al.(2014)]{Kraus}Kraus, M., Cidale, L. S., Arias, M. L., 
    Oksala, M. E. \& Borges Fernandes, M. 2014, \apjl, 780, L10
%%  
\bibitem[Larsen \& Richtler(1999)]{Larsen99}Larsen, S. S. \& Richtler, T. 1999, \aap, 345, 59
%%  
\bibitem[Makovoz \& Khan(2005)]{MK2005}Makovoz, D. \& Khan, I. 2005, Astronomical 
  Data Analysis Software and Systems XIV (ASP Conf. Ser. 347), ed. P. Shopbell, 
  M. Britton, \& R. Ebert (San Francisco, CA: ASP), 81 
%%  
\bibitem[Makovoz \& Marleau(2005)]{MM2005}Makovoz, D. \& Marleau, F. R. 2005, 
   PASP, 117, 1113
%%  
\bibitem[Martins et al.(2005)]{Martins}Martins, F., Schaerer, D. \& Hillier, D. J. 
  2005, \aap, 436, 1049 
%%   
\bibitem[Margutti, et al.(2014)]{Margutti}Margutti, R. et al. \apj, 780, 21
%%  
\bibitem[Mauerhan, et al.(2013a)]{Mauerhan}Mauerhan, J. C., Smith, N.,, Filippenko, A. V., et al 2013a, \mnras, 430, 181 
%%  
\bibitem[Mauerhan, et al.(2013b)]{Mauerhanb}Mauerhan, J. C., Smith, N., Silverman, J.M. et al. 2013b, \mnras, 431, 2599  
\bibitem[Maund et al.(2006)]{Maund06}Maund, J. R., et al. 2006, \mnras, 369, 390 
%%  
\bibitem[Maund et al.(2008)]{Maund08}Maund, J. R., et al. 2008, \mnras, 387, 1344
%%  
\bibitem[Mehner et al.(2013)]{Mehner}Mehner, A. et al. 2013, \aap, 555, A116 
%%  
\bibitem[Oksala et al.(2013)]{Oksala}Oksala, M. E., Kraus, M., Cidale, L. S., 
    Muratore, M. F., \& Borges Fernandes, 2013, \aap, 558, A17 
    %%
\bibitem[Owocki \& Shaviv(2016)]{Owocki}Owocki, S. P. \&  Shaviv, N. J. 2016, \mnras,
 462, 345 
%% 
\bibitem[Pastorello et al.(2010)]{Pastorello}Pastorello,A., et al. 2010, \mnras, 408,
 181 
%%  
\bibitem[Pastorello et al.(2013)]{Pastorello13}Pastorello,A., et al. 2013, \apj, 767, 1 
%%  
\bibitem[Schlafly \& Finkbeiner(2011)]{SF2011}Schlafly, E. F. \& Finkbeiner, D. P. 
 2011, \apj, 737, 103 
%%  
\bibitem[Schwarz \& Li(2003)]{Schwarz03}Schwarz, M. \& Li, W. 2003, IAU Circ. 8051 
  %%  
\bibitem[Sirianni et al.(2005)]{Sirianni}Sirianni, M. et al. 2005, \pasp, 117, 1049 
 %%  
 \bibitem[Smith et al.(2001)]{Smith}Smith, N., Humphreys, R. M., \& Gerhrz, R. D. 2001 , \pasp, 113, 692 
%%  
\bibitem[Smith et al.(2010)]{Smith2010}Smith, N. et al. 2010 \aj, 139, 1451
%%
\bibitem[Smith \& Tombleson(2015)]{Smith15}Smith, N. \& Tombleson, R. 2015, 
\mnras, 447, 598 
%%
\bibitem[Smith(2016)]{Smith2016}Smith, N. 2016, \mnras, 461, 3353 
%%  
\bibitem[Suh(1999)]{Suh}Suh, K.-W. 1999, \mnras, 304, 389
%%
\bibitem[Tammann \& Sandage(1968)]{ST1968}Tammann,G. A. \& Sandage, A. 1968, 
    \apj, 151, 825
%%
\bibitem[Th\'{e} \& Graaland(1995)]{The}Th\'{e}, P. S. \& Graaland, F. 1995, in Rev. Mex. A \& A, Conf. Sef. 2, 75
%%  
\bibitem[Van Dyk et al.(2005a)]{VanDyk05a}Van Dyk, S. D. 2005a, in The Fate of 
  the Most Massive Stars, ed. R. Humphreys \& K. Stanek (San Francisco:ASP), 49
%%  
\bibitem[Van Dyk et al.(2005b)]{VanDyk05b}Van Dyk, S. D., Filippenko, A. V., 
  Chornock, R., Li, Weidong, \& Challis, P. M. 2005b, \pasp, 117, 553 
%%  
 \bibitem[Van Dyk et al.(2006)]{VanDyk06}Van Dyk, S. D, Li, Weidong, Filippenko, A. V., Humphreys, R. M., Chornock, R.,, Foley, R., \& Challis, P. M. 2006, asXiv:astro-ph/0603025v1
%%  
\bibitem[Van Dyk \& Matheson(2012a)]{VanDyk12a}Van Dyk, S. D. \&  Matheson, T. 2012a, \apj,746, 179
%%  
\bibitem[Van Dyk \&  Matheson(2012b)]{VanDyk12b}Van Dyk, S. D. \& Matheson, T. 2012b, in Eta Carinae  and the Supernova Impostors, Astrophys.\ \& Sp.\ Sci.\ Library 384 (ed .\ K.\ Davidson \& R.M.\ Humphreys, Springer Media, New York),  249 
%%  
\bibitem[Vink(2012)]{Vink2012}Vink, J. S. 2012, in Eta Carinae  and the Supernova Impo
stors, Astrophys.\ \& Sp.\ Sci.\ Library 384 (ed.\ K.\ Davidson \& R.M.\ Humph
reys, Springer Media, New York), 221
%% 
\bibitem[Wagner et al.(2004)]{Wagner}Wagner, R. M., et al. 2004, PASP, 116, 326 
%%  
\bibitem[Weis \& Bomans(2005)]{Weis05}Weis, K. \& Bomans, D.~J. 2005, \aap, 429, L13
%%  
\bibitem[Wolf, Appenzeller, \& Stahl(1981)]{Wolf81}Wolf, B., Appenzeller, I., 
  \& Stahl, O. 1981, \aap, 103, 94 
%%  
\bibitem[Wolf(1989)]{Wolf}Wolf, B. 1989, \aap, 217, 87 
%%  
\end{thebibliography}
\end{document}